\documentclass{ectj}

\usepackage{amsfonts,amssymb,graphics,epsfig,verbatim,bm,latexsym,amsmath,url,amsbsy}

\newtheorem{theorem}{Theorem}

\newtheorem{corollary}{Corollary}

\newtheorem{definition}{Definition}
\newtheorem{rules}{Rule}

\renewcommand{\thesection}{\arabic{section}}
\renewcommand{\theequation}{\arabic{section}.\arabic{equation}}

\newcounter{bean}

\received{September 2021}
\accepted{February 2023}
\volume{20}

\usepackage{tikz}
	\usetikzlibrary{arrows,positioning} 
\usepackage{subcaption}
\usepackage{bbm}

\newcommand\independent{\protect\mathpalette{\protect\independenT}{\perp}}
\def\independenT#1#2{\mathrel{\rlap{$#1#2$}\mkern2mu{#1#2}}}

\DeclareFontFamily{OT1}{pzc}{}
\DeclareFontShape{OT1}{pzc}{m}{it}{<-> s * [1.10] pzcmi7t}{}
\DeclareMathAlphabet{\mathpzc}{OT1}{pzc}{m}{it}

\title[Causal Inference and Data Fusion in Econometrics]{Causal Inference and Data Fusion in Econometrics}

\author[P.\ H\"{u}nermund and B.\ Bareinboim]{Paul H\"{u}nermund$^{\dagger}$ and
                Elias Bareinboim$^{\ddagger}$}

\address{$^{\dagger}$Copenhagen Business School, 
									 Kilevej 14A, Frederiksberg, 
									 2000, DK}
\email{phu.si@cbs.dk}

\address{$^{\ddagger}$Columbia University, 
									   500 W 120th Street, New York,
									   NY 10027, USA}
\email{eb@cs.columbia.edu}

\def\AmSTeX{$\cal A$\kern-.1667em\lower.5ex\hbox{$\cal M$}\kern-.125em
    $\cal S$-\TeX}
\def\BibTeX{{\rm B\kern-.05em{\sc i\kern-.025em b}\kern-.08em
    T\kern-.1667em\lower.7ex\hbox{E}\kern-.125emX}}

\begin{document}

    \begin{abstract}
		Learning about cause and effect is arguably the main goal in applied econometrics. In practice, the validity of these  causal inferences is contingent on a number of critical assumptions regarding the type of data that has been collected and the substantive knowledge that is available about the phenomenon under investigation. For instance, unobserved confounding factors threaten the internal validity of estimates; data availability is often limited to non-random, selection-biased samples; causal effects need to be learned from surrogate experiments with imperfect compliance; and causal knowledge has to be extrapolated across structurally heterogeneous populations. A powerful and flexible causal inference framework is required in order to tackle all of these challenges, which plague essentially any data analysis to varying degrees. Building on the structural perspective on causality introduced by \citet{Haavelmo1943} and the graph-theoretic approach proposed by \citet{Pearl1995}, the artificial intelligence (AI) literature has developed a wide array of techniques for causal inference that allow to leverage information from various imperfect, heterogeneous, and biased data sources (\citealp{Bareinboim2016}). In this paper, we review recent advances made in this literature that have the potential to contribute to econometric methodology along three broad dimensions. First, they provide a unified and comprehensive framework for causal learning, in which the above-mentioned problems can be addressed in generality. Second, due to their origin in AI, they come together with sound, efficient, and complete (to be formally defined) algorithmic criteria for automation of the corresponding identification task. And third, because of the nonparametric description of structural models that graph-theoretic approaches build on, they combine the analytical rigor of structural econometrics with the flexibility of the potential outcomes framework, and thus offer a valuable complement to these two literature streams.


    \end{abstract}

    % main body

   %%%%%%%%%%%%%%%%%%%%%%%%%%%%%%%%%%%%%%%%%%%%%%
\section{Introduction}
\label{sec_introduction}
%%%%%%%%%%%%%%%%%%%%%%%%%%%%%%%%%%%%%%%%%%%%%%

\setcounter{equation}{0}

Obtaining causal knowledge by uncovering quantitative relationships in statistical data is arguably one of the most important goals of econometrics since the beginning of the discipline. Policy-makers, legislators, and managers need to be able to forecast the likely impact of their actions in order to make informed decisions. Phillip G.\ Wright's \citeyearpar{Wright1928} seminal contribution on instrumental variable estimation, using the theory of path coefficients developed by his son Sewall \citet{Wright1921,Wright1923}, for example, was motivated by the desire to understand the effect of tariffs on the production and import of agricultural products. In the postwar period, interest in the topic of causal inference initially experienced a decline in attention (\citealp{Hoover2004}), but was brought back to the forefront of the methodological debate by the emergence of the potential outcomes framework (\citealp{Rubin1974,Imbens2015,Imbens2019}) and advances in structural econometrics (\citealp{Heckman2007, Matzkin2013,Lewbel2019}).

\citet{Woodward2003} defines causal knowledge as ``knowledge that is useful for a very specific kind of prediction problem: the problem an actor faces when she must predict what would happen if she or some other agent were to act in a certain way [...]".\footnote{Woodward continues: ``[...] on the basis of observations of situations in which she or the other agent have not (yet) acted" (p.\ 32).} This association of causation with control in a stimulus-response-type relationship is likewise foundational for econometric methodology. Following \citet{Strotz1960}, ``$z$ is a cause of $y$ if [...] it is or 'would be' possible by \emph{controlling} $z$ indirectly to control $y$, at least stochastically" (p.\ 418; emphasis in original). 

Although implicit in earlier treatments in the field (e.g., \citealp{Haavelmo1943}), \citet{Strotz1960} were the first to express actions and control of variables as \emph{``wiping out"} of structural equations in an economic system (\citealp{Pearl2009}, p.\ 32). To illustrate this idea, consider the two-equation model
\begin{align}
Z &= f_z (W, U_z), \\[5pt]
Y &= f_y (Z, W, U_y),
\end{align}
in which $Y$ represents the outcome of interest, $Z$ a treatment under study, $W$ other socioeconomic variables, and $U$ unobserved exogenous background factors.\footnote{We follow the usual notation of denoting random variables by uppercase and their realized values by lowercase letters.} Since $W$ enters in both equations of the system, it creates a correlation between $Z$ and $Y$ that is not the result of a causal impact. Therefore, to predict how $Y$ reacts to induced changes in $Z$, the causal mechanism that naturally determines $Z$ needs to be replaced to avoid non-causal (spurious) sources of variation. In this  particular example, the values that $Z$ attains must be decoupled from $W$, so that $Z$ can freely influence $Y$. Symbolically, this is achieved by deleting $f_z(\cdot)$ from the model and fixing $Z$ at a constant value $z_0$. The modified system thus becomes
\begin{align}
Z &= z_0, \tag{1.1'} \\[5pt]
Y &= f_y (z_0, W, U_y). \tag{1.2'} \label{eq1_2_prime}
\end{align}
Subsequently, the quantitative impact of the intervention can be traced via equation (\ref{eq1_2_prime}) to pin down $Z$'s  causal effect on $Y$.

The notion of wiping out equations, as proposed by Strotz and Wold, eventually received central status and a formal treatment in a specific language with the definition of the \emph{do}-operator (\citealp{Pearl1995}). Consider the task of predicting the post-intervention distribution of a random variable $Y$ that is the result of a manipulation of another variable $X$. In mathematical notation, this can be written as $Q=P(Y=y|do(X=x))$, where $do(X=x)$ denotes the replacement of whatever mechanisms were there for $X$, say $f_x$, by a constant $x$.  

In practical applications, however, simulating interventions to such a degree of granularity would either require knowledge about the precise form of the system's underlying causal mechanisms or the possibility to physically manipulate $X$ in a controlled experiment. Both are luxuries that policy forecasters often do not have available. In many economic settings, experiments can be difficult to implement, due to cost, technical, or ethical considerations. Likewise, exactly knowing the structural mechanisms that truly govern the data generating process is hard in the social sciences, where often only qualitative knowledge about causal relationships is available.\footnote{Quoting prominent physicist Murray Gell-Mann: \emph{``Imagine how hard physics would be if electrons could think.''} (cited in \citealp{Page1999}).} This means that the counterfactual distribution $Q = P(y|do(x))$ will, in general, not be immediately estimable. In practice, instead, $Q$ will first need to be transformed into a standard probability object that only comprises ex-post observable quantities before estimation can proceed. The symbolic machinery that warrants such kinds of syntactic transformations is called \emph{do-calculus} (\citealp{Pearl1995}).

\begin{figure}[t]
	\includegraphics[width=\textwidth]{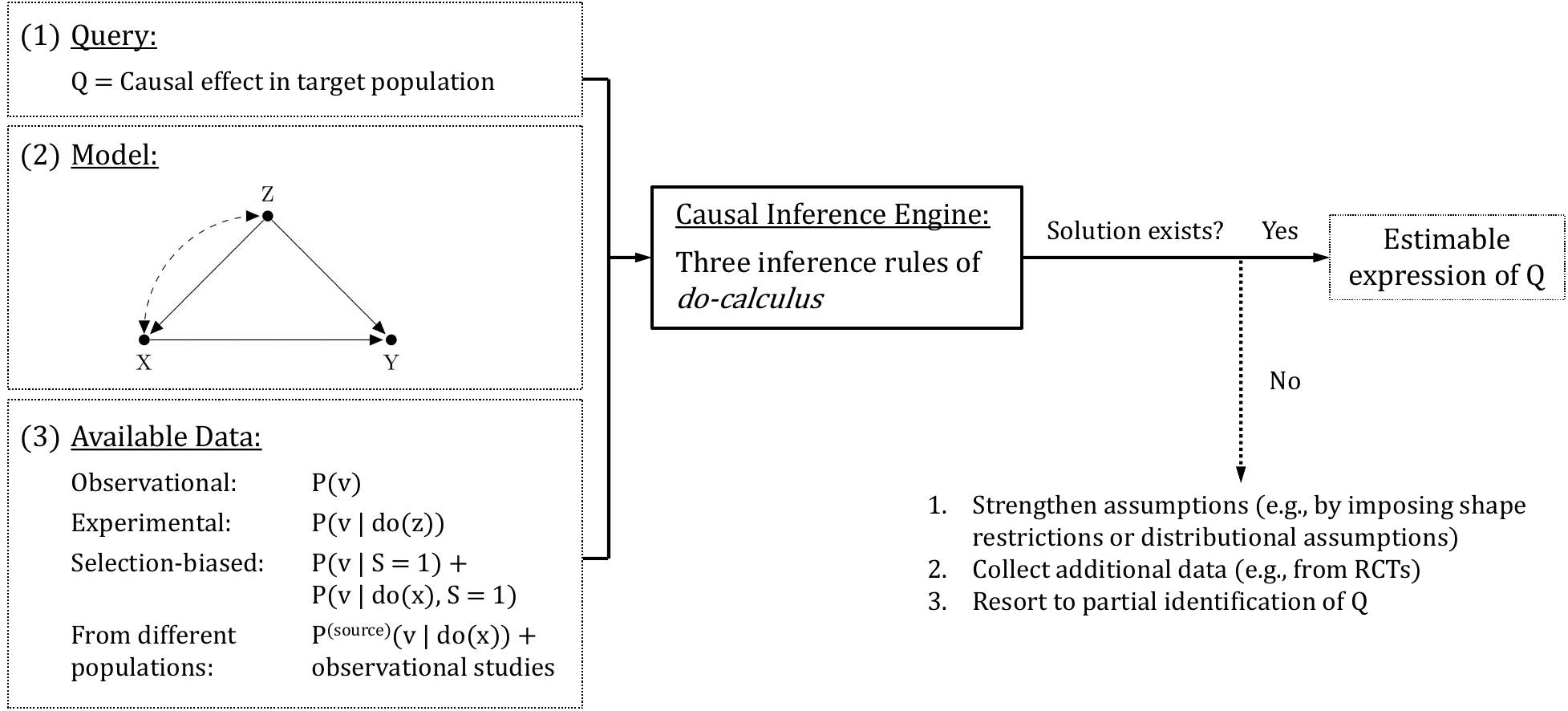}
	\caption{Schematic illustration of the data fusion process.}
	\label{fig_causal_inference_engine}
\end{figure}

Do-calculus can be seen as a causal inference engine that takes three inputs:
\setcounter{bean}{0}
\begin{list}
	{\arabic{bean}.}{\usecounter{bean}}
	 \item A causal quantity $Q$, which is the query the researcher wants to answer; 
	 \item A model $G$ that encodes the qualitative understanding about the structural dependencies between the economic variables under study; and 
	 \item  A collection of datasets $P(v|\cdot)$ that are available to the analyst, including  observational, experimental, selection-biased samples, from different populations, and so on.
\end{list} 
Building on these inputs, do-calculus consists of three inference rules for transforming probabilistic sentences involving do-expressions into equivalent expressions. The inferential goal is then to re-express the causal quantity $Q$ (1.\ above) through the repeated application of the rules of the calculus, licensed by the assumptions in $G$ (2.\ above), into expressions that are estimable from the observable probability distributions $P(v|\cdot)$ (3.\ above). Figure \ref{fig_causal_inference_engine} provides a schematic illustration of this process.

Do-calculus complements standard tools in econometrics in two important ways. First, it builds on a mathematical formalism borrowed from graph theory, which describes causal models as a set of nodes in a network, connected by directed edges (so-called \emph{directed acyclic graphs}; \citealp{Pearl1995}). An advantage of such a description is that it does not rely on any functional-form restrictions imposed on the relationships between economic variables. Therefore, the approach provides a fully general, formal treatment of nonparametric causal inference (i.e., relying solely on exclusion and independence restrictions encoded in the graph) in recursive models. Second, as a subfield of artificial intelligence, the literature on graph-theoretic treatments of causality has developed algorithmic solutions for a wide variety of causal inference problems arising in applied work. These algorithms are able to carry out the syntactic transformation described above -- mapping a query to the available data through the model's assumptions -- fully automatically. From do-calculus, the algorithms furthermore inherit the property of \emph{soundness} and \emph{completeness} (\citealp{Tian2002,Shpitser2006,Huang2006,Bareinboim2012b,Lee:etal19,NEURIPS2020_7b497aa1,Correa2021}). This means that the approach is guaranteed to return a correct solution whenever one exists. Conversely, and remarkably, if the algorithm fails to provide an answer to a causal query, it is assured that no such answer will be obtainable unless the assumptions imposed on the model are strengthened. In other words, for the class of models in which these algorithmic conditions are applicable, the identification problem is fully solved (\citealp{Pearl2013, Bareinboim2016}).\footnote{These tools are implemented in the free software \url{www.causalfusion.net} and available to researchers.}

The development of do-calculus gave the literature on causal inference within the field of artificial intelligence a tremendous boost, and many significant advances have been made since \citet{Pearl2000} published his seminal contribution. The aim of this paper is to discuss these more recent developments and show how do-calculus  can be utilized to solve many recurrent problems in applied econometric work.\footnote{There is a growing interest in the graph-theoretic approach to causal inference in economics. Early examples constitute  \citet{Adams2003}, \citet{Neuberg2003}, \citet{White2006}, and \citet{Froelich2008}, and more recent ones include \citet{Heckman2013} (examined in detail in \citealp{Pearl2013}) and \citet{Imbens2019}. However, these papers do not cover the newer advances that were made in the causal AI literature in the last decade, which is the primary focus of this paper.} 
The three main topics we cover are: dealing with confounding bias (Section \ref{sec_confounding}), recovering from sample selection bias (Section \ref{sec_selection}), and extrapolation of causal claims across heterogeneous settings (Section \ref{sec_transportability}), which we describe in turn next.

\emph{Confounding bias (Section \ref{sec_confounding})}. In most applied settings, the post-interventional distribution of $Y$ following a manipulation of $X$, $P(y|do(x))$, does not coincide with the conditional, observational distribution $P(y|x)$ -- a distinction that has been popularized through the mantra \emph{``correlation does not imply causation"}. This is due to confounding influence factors, which can render two variables stochastically dependent irrespective of any causal relationship between them. The inference rules of do-calculus were developed precisely to neutralize confounding bias. Syntactically, this task amounts to transforming $P(y|do(x))$ into an equivalent expression, generally different from $P(y|x)$, that is nonetheless estimable from the available data. If a reduction containing standard probability objects can be reached, the confounding problem is solvable with the help of observational data alone. Additionally, sometimes the analyst is able to experimentally manipulate a third variable $Z$, which is itself causally related to the treatment of interest. In such settings, the identification problem can be relaxed, since estimable syntactic transformations of $P(y|do(x))$ reached by do-calculus can now also involve $do(z)$-distributions.

\emph{Sample selection bias (Section \ref{sec_selection})}. A common threat to the validity of  inferences in practice is sample selection bias, which occurs if the analyst is only able to observe information for members of the population that possess specific characteristics or fulfil certain requirements  (e.g., market wages are only observable if individuals are employed; \citealp{Heckman1979}). Selection-biased data aggravate the identification problem as $P(y|do(x))$ needs to be transformed into an expression solely comprised of probabilities from a non-random sample of the population (inclusion in the selected sample is usually denoted by an indicator $S$, which implies that only probabilities conditional on $S=1$ are observable). The inference rules of do-calculus provide a principled and complete solution for carrying out this task.

\emph{Extrapolation of causal claims across settings (Section \ref{sec_transportability})}. While confounding and selection biases threaten the internal validity of estimates, another important topic in econometric practice is external validity, or generalizability of causal inferences across settings and populations. Causal knowledge is usually acquired in a specific population (e.g., for subjects in a laboratory setting), but needs to be brought to productive use in other domains to be most valuable. What permits such a transportation of causal knowledge across settings, however, if the underlying populations differ structurally in important ways? Do-calculus provides an answer to this question. Its inference rules can be applied to transform a causal query in a target population into an expression that is estimable with the help of information stemming from a source population. In its more general form, transportability theory encompasses the problem of combining causal knowledge from several, possibly heterogeneous source domains (a strategy generically known under the rubric of \emph{``meta-analysis"}). Thereby, do-calculus opens up entirely new possibilities for leveraging results from a whole body of empirical literature to address policy questions arising in yet under-researched contexts.

These three thematic areas are indeed quite diverse and encompass several seemingly unrelated empirical challenges; yet, they share a common structure. Data, which are created in various different ways -- e.g., from observational or experimental studies, from non-random sampling, or from heterogeneous underlying populations -- are combined in order to answer a causal query of interest. For this strategy of \emph{``data fusion"} (see Figure \ref{fig_causal_inference_engine}) to be viable, the analyst needs to be equipped with a model of the underlying economic context under study and a powerful inference framework that licenses this kind of information transfer and reconciliation (\citealp{Bareinboim2016}). In the remainder of the paper, we will describe such a causal modelling and inference framework in detail.  

   %%%%%%%%%%%%%%%%%%%%%%%%%%%%%%%%%%%%%%%%%%%%%%
\section{Preliminaries: Structural Causal Models, Causal Graphs, and Interventions}
\label{sec_preliminaries}
%%%%%%%%%%%%%%%%%%%%%%%%%%%%%%%%%%%%%%%%%%%%%%

\setcounter{equation}{0}

This section introduces structural causal models (SCM) and directed acyclic graphs, which form the basis for all the data fusion techniques discussed in this paper.\footnote{Structural causal models are nonparametric versions of structural equation models (SEM). We purposefully will use the term SCM to avoid confusion with the vast literature on SEM that traditionally assumes parametric or even linear functional forms, and many times has confounded the inherent causal nature of structural models.} We follow the standard notation in the literature, as summarized in \citet{Pearl2009}, and define an SCM as

\begin{definition}
	\label{def_scm}
	(Structural causal model; \citealp{Pearl2009}) A structural causal model is a 4-tuple $M = \langle U, V, F, P(u) \rangle$ where
	
	\setcounter{bean}{0}
	\begin{list}
	{\arabic{bean}.}{\usecounter{bean}}
	\item $U$ is a set of background variables (also called exogenous) that are determined by factors outside the model.
	\item $V = \{V_1,\hdots,V_n\}$ is a set of endogenous variables that are determined by variables in the model, viz.\ variables in $U \cup V$.
	\item $F$ is a set of functions $\{f_1,\hdots, f_n\}$ such that each $f_i$ is a mapping from (the respective domains of) $U_i \cup PA_i$ to $V_i$, where $U_i \subseteq U$ and $PA_i \subseteq V \setminus V_i$ and the entire set of $F$ forms a mapping from $U$ to $V$. In other words, $f_i$ assigns a value to the corresponding $V_i \in V$,  $v_i \leftarrow f_i(pa_i, u_i)$, for $i = 1,\hdots, n$. 
	\item $P(u)$ is a probability function defined over the domain of $U$.
	\end{list}
\end{definition}

An SCM constitutes a set of (exogenous) background factors, $U$, which are determined outside of the model. Their associated (joint) probability distribution, $P(u)$, creates variation in the endogenous variables, $V$, whose source remains not further specified. Inside the model, the value of an endogenous variable $V_i$ is determined by a causal process, $v_i \leftarrow f_i(pa_i, u_i)$, that maps the background factors $U_i$ and a set of endogenous variables $PA_i$ (so-called \emph{parents}) into $V_i$. These causal processes -- or mechanisms -- are assumed to be invariant unless explicitly intervened on (see Section \ref{sec_interventions}). Together with the background factors, they represent the data generating process (DGP) according to which \emph{nature} assigns values to the (endogenous) variables under study.\footnote{Background factors correspond to what is often referred to as ``error terms" in classical econometrics. However, we deliberately avoid this terminology to emphasize that the $U_i$'s in an SCM have a causal interpretation, in contrast to the purely statistical notion of a prediction error or deviation from a conditional mean function.}

To emphasize the interpretation of $f_i$'s as stimulus-response relationships, and in contrast to the standard notation in econometrics, the artificial intelligence literature uses assignment operators ``$\leftarrow$" instead of equality signs (similar to the syntax of programming languages). Assignments change meaning under solution-preserving algebraic operations; i.e., $y \leftarrow ax \neq x \leftarrow y/a$ (\citealp{Pearl2009}, p.\ 27). This highlights the asymmetric nature of elementary causal mechanisms (\citealp{Woodward2003,Cartwright2007}), in the sense that if $X$ is a cause of $Y$, it cannot be the case that $Y$ is also a cause of $X$ at the same instance of time.

In a fully specified SCM, $\langle U, V, F, P(u) \rangle$, any counterfactual quantity is well-defined and immediately computable from the model. In many social science contexts, however, precise knowledge of the functional relationships, $f_i$, and the distribution of the background factors, $P(u)$, governing the DGP, is not available. In the following, we will thus advocate for an approach that fully embraces and acknowledges the existence of the underlying causal mechanisms and exogenous variations in the system (i.e., nature follows a structural causal model), but which will be much less committal regarding what the analyst needs to know about this reality in order to be able to make causal inferences.  In particular, the inferences entailed by our analysis will rely on the graphical representation of the underlying structural system, which is a way of encoding a parsimonious set of assumptions of the system sufficient for identifiability.

\begin{figure}[thp]
	\centering
	\subcaptionbox{\label{fig1a}}[.36\linewidth]{
		\begin{tikzpicture}[>=triangle 45, font=\footnotesize]
		% Nodes
		\node[fill,circle,inner sep=0pt,minimum size=5pt,label={below:{X}}] (X) at (0,0) {};
		\node[fill,circle,inner sep=0pt,minimum size=5pt,label={right:{Z}}] (Z) at (1.5,1.5) {};
		\node[fill,circle,inner sep=0pt,minimum size=5pt,label={below:{Y}}] (Y) at (3,0) {};
		\node[fill,circle,inner sep=0pt,minimum size=5pt,label={above:{U$_X$}}] (UX) at (-0.7,0.7) {};
		\node[fill,circle,inner sep=0pt,minimum size=5pt,label={above:{U$_Z$}}] (UZ) at (1.5,2.49) {};
		\node[fill,circle,inner sep=0pt,minimum size=5pt,label={above:{U$_Y$}}] (UY) at (3.7,0.7) {};
		% Arcs
		\draw[->,shorten >= 1pt] (Z)--(X);
		\draw[->,shorten >= 1pt] (Z)--(Y);
		\draw[->,shorten >= 1pt] (X)--(Y);
		\draw[->,dashed,shorten >= 1pt] (UX)--(X);
		\draw[->,dashed,shorten >= 1pt] (UZ)--(Z);
		\draw[->,dashed,shorten >= 1pt] (UY)--(Y);
		\end{tikzpicture}
	}
	\quad
	\subcaptionbox{\label{d-separation}}[.24\linewidth]{
		\begin{tikzpicture}[>=triangle 45, font=\footnotesize]
		\node[fill,circle,inner sep=0pt,minimum size=5pt,label={below:{A}}] (A) at (0,2) {};
		\node[fill,circle,inner sep=0pt,minimum size=5pt,label={[label distance=0pt,xshift=-0.1cm,yshift=-0.15cm]right:{D}}] (D) at (1,1) {};
		\node[fill,circle,inner sep=0pt,minimum size=5pt,label={below:{B}}] (B) at (2,2) {};
		\node[fill,circle,inner sep=0pt,minimum size=5pt,label={below:{C}}] (C) at (3.4,2) {};
		\node[fill,circle,inner sep=0pt,minimum size=5pt,label={below:{E}}] (E) at (1,0) {};
		\draw[->,shorten >= 1pt] (A)--(D);
		\draw[->,shorten >= 1pt] (B)--(C);
		\draw[->,shorten >= 1pt] (B)--(D);
		\draw[->,shorten >= 1pt] (D)--(E);
		\end{tikzpicture}
	} 
	\qquad
	\subcaptionbox{\label{fig1c}}[.28\linewidth]{
		\begin{tikzpicture}[>=triangle 45, font=\footnotesize]
		% Nodes
		\node[fill,circle,inner sep=0pt,minimum size=5pt,label={below:{X}}] (X) at (0,0) {};
		\node[fill,circle,inner sep=0pt,minimum size=5pt,label={above:{Z}}] (Z) at (1.5,1.5) {};
		\node[fill,circle,inner sep=0pt,minimum size=5pt,label={below:{Y}}] (Y) at (3,0) {};
		\node[fill,circle,inner sep=0pt,minimum size=5pt,label={above:{x$_0$}}] (x0) at (0,0.8) {};
		% Arcs
		\draw[->,shorten >= 1pt] (Z)--(Y);
		\draw[->,shorten >= 1pt] (X)--(Y);
		\draw[->,shorten >= 1pt] (x0)--(X);
		\end{tikzpicture}
	}
	\caption{Examples of directed acyclic graphs representing structural causal models.}
	\label{fig1}
\end{figure}

Every SCM $M$ defines a directed graph $G(M)$ (or $G$, for simplicity). Nodes in $G$ correspond to endogenous variables in $V$, and directed edges point from the set of parent nodes $PA_i$ towards $V_i$.\footnote{As it is common in graph theory, we will use the notation of kinship relations (parents, children, ancestors, descendants, etc.) to describe the relative position of nodes in directed graphs. For instance, for the graph in Figure \ref{d-separation} we can read that $B$ is a parent of $D$, since $B \rightarrow D$, $A$ is an ancestor of $E$, since $A \rightarrow D \rightarrow E$, and $E$ is a child of $D$, since $D \rightarrow E$.} An example is given in Figure \ref{fig1a}, which refers to the following underlying structural causal model
\begin{equation}
\label{eq2_1}
\begin{aligned}
Z &\leftarrow f_z(U_z), \\
X &\leftarrow f_x(Z, U_x), \\
Y &\leftarrow f_y(X, Z, U_y).
\end{aligned}
\end{equation}
Note that $Z$ appears as an argument in the structural function of $X$, $f_x$. Accordingly, $Z$ is a parent of $X$ and an arrow should be added pointing from node $Z$ to $X$. Similarly, $X$ and $Z$ appear in $f_y$, which means that the causal graph contains arrows from these variables to $Y$. For the sake of readability, we will usually not depict the $U_i$'s explicitly, as in Figure \ref{fig1a}, but will omit them from the graph whenever they affect only one endogenous variable at a time. The presence of common unobserved parent nodes, which render two variables stochastically dependent, is represented by dashed bidirected arcs in the graph (see, e.g., Figure \ref{fig_college}). I.e., the arc $X \dashleftarrow \dashrightarrow Y$ serves as a shortcut notation for $X \leftarrow U \rightarrow Y$, where the set of common causes $U$ is unobservable to the analyst.

The graph in Figure \ref{fig1a} contains no sequences of edges that point from a variable back to itself (i.e., there are no feedback loops). This property is called \emph{acyclicity}. Throughout the paper, we restrict attention to structural causal models that can be represented by directed acyclic graphs (DAG). This class of models, which economists refer to as \emph{recursive}, is of central importance in causal inference, because they describe economic systems in which individual causal mechanisms have a direct and autonomous stimulus-response interpretation, in accordance with the notion of causality put forward by \citeauthor{Strotz1960} (1960; see also \citealp{Woodward2003, Cartwright2007}).\footnote{Incidentally, the potential outcomes framework in the econometric treatment effects literature also interprets the link between treatment and outcome as a stimulus-response relationship and therefore implicitly maintains the assumption of acyclicty (\citealp{Heckman2007}). To witness the discussion about the causal interpretation of individual functional relationships in recursive versus nonrecursive models in the early econometrics literature, see \citet{Haavelmo1943}, \citeauthor{Bentzel1946} (1946; as cited in \citealp{Wold1981}), \citet{Bentzel1954}, \citet{Strotz1960}, \citet{Wold1960}, and \citet{Basman1963}.} It is important to note, however, that the axioms of structural counterfactuals in SCMs (\citealp{Pearl2009}, ch.\ 7) also hold in nonrecursive models, as discussed in \citet{halpern:2000}. Graphical causal model with directed cycles, which can incorporate feedback loops and equilibrium behavior, are an active area of research. The interested reader is referred to \citet[][ch.\ 12]{Spirtes2001}, \citet[][ch.\ 3.6]{Pearl2009}, and \citet{Bongers2021}.

Working with the graphical representation of an SCM entails a deliberate choice by the analyst to refrain from distributional and functional form assumptions, since the shape of the $f_i$'s and the distribution of background factors $U_i$ remain unspecified throughout the analysis. Another way of thinking about the causal graph is that it represents the equivalence class of all structural functions sharing the same scope. Consequently, graphical models are fully nonparametric in nature.\footnote{Note that ``nonparametric" in the artificial intelligence literature refers to the absence of assumptions involving error distributions \emph{as well as} constraints over the form of the structural functions in the SCM. Instead, the shared features assumed to be available across structural systems are topological, that is, exclusion and independence restrictions are encoded in the causal graph. This difference in terminology should be kept in mind for what follows.} Shape restrictions on functions (such as separability, monotonicity, or differentiability) and distributional assumptions might sometimes be licensed by economic theory (\citealp{Heckman2007,Matzkin2007,Matzkin2013}). In case they are not, however, we concur with \citet{Manski2003} that it is a more robust research approach to start with the most flexible model possible and only resort to distributional and functional form assumptions once the explanatory power of nonparametric identification approaches has been exhausted. In line with this philosophy, the techniques we present in the following explore ways to identify causal effects from data when only knowledge about the graph $G$ is available.\footnote{This is indeed the case unless otherwise specified, and should constitute the starting point of any analysis. Whenever nonparametric identification is not entailed by  the available knowledge, the causal graph can still be used as a computation device to analyze identifiability of entire classes of structural models. For instance, the most general identification results of structural coefficients if the system is linear are within the graphical perspective. For a survey and the latest results, please refer to \citet[ch.\ 5]{Pearl2009} and
\citet{kumor2020auxiliary}.} 
%chen:etal17,

One key property of DAGs is that they are falsifiable through testable implications over the observed distributions, including conditional independence relationships between variables in the model.\footnote{Historically, DAGs were first introduced to the AI literature in the early 1980s as efficient encoders of conditional independence constraints, and as a basis that avoided the explicit enumeration of exponentially many of such constraints. This encoding lead to a huge literature on efficient algorithms for computing and updating probabilistic relationships in data-intensive applications (\citealp{Pearl1988}).} We define below such notion. 

\begin{definition}
	\label{def_d-separation}
	(D-separation; \citealp{Pearl1988}) A set $Z$ of nodes is said to block a path $p$ if either: (a) $p$ contains at least one arrow-emitting node that is in $Z$, or (b) $p$ contains at least one collision node that is outside $Z$ and has no descendant in $Z$. If $Z$ blocks \emph{all} paths from set $X$ to set $Y$, it is said to ``d-separate $X$ and $Y$", and then it can be shown that variables $X$ and $Y$ are independent given $Z$, written as $X \independent Y | Z$.\footnote{See \citet{Verma1988}. A path refers to any consecutive sequence of edges in a graph. The orientation of edges plays no role. If the direction of edges is taken into account, one speaks of a \emph{directed} or \emph{causal path}: $A \rightarrow B \rightarrow C$.}
\end{definition}

Conditional independence licensed by d-separation (\emph{d} stands for ``directional") holds for any distribution $P(v)$ over the variables in the model that is compatible with the causal assumptions encoded in the graph. Remarkably, this is true \emph{regardless} of the parametrization of the arrows. An example is given in Figure \ref{d-separation}, where the path $A \rightarrow D \leftarrow B \rightarrow C$ is blocked by $Z = \{B\}$, since $B$ emits arrows on that path. Consequently, we can infer the conditional independencies $A \independent C | B$ and $D \independent C | B$. In fact, $A$ and $C$ are independent conditional on the empty set $\{\emptyset\}$ as well. $D$ acts as a so-called \emph{collider} node in this path, because of two arrows pointing into it. Therefore, according to the second condition of Definition \ref{def_d-separation}, the path between $A$ and $C$ is blocked without any conditioning. Conversely, when conditioned on, a collider would open up a path that has been previously blocked; thus, $A \not\!\perp\!\!\!\!\perp C | D$. The same holds for descendants of colliders such as $E$ in Figure \ref{d-separation}, yielding $A \not\!\perp\!\!\!\!\perp C | E$.

D-separation allows to systematically read off the conditional independencies implied by the structural model from the graph.\footnote{\citet{Anand2021} prove d-separation as well as further identification results for so-called \emph{cluster causal diagrams} (C-DAGs), which allow for the partial specification of relationships among variables based on limited prior knowledge.} As mentioned earlier, this method provides the analyst with a set of testable implications that can be benchmarked with the available data. The full list of conditional independence relations (with separator sets up to cardinality one) implied by the graph in Figure \ref{d-separation} is given by
\begin{gather}
\begin{split}
A \independent B; \qquad A \independent C; \qquad A \independent E | D; \qquad B \independent E | D; \\
C \independent D | B; \qquad C \independent E | D; \qquad C \independent E | B. 
\end{split}
\end{gather}
These independence relations can be checked using statistical hypothesis testing, and if rejected, the hypothesized model could be refuted and/or revised. An advantage of such local tests, compared to global goodness-of-fit measures, for example, is that they indicate exactly where the model is incompatible with the observed data. Thus, the analyst can rely on concrete clues about where to improve the model, which facilitates an iterative process of model building and criticism.

Conditional independence assumptions are one of the main building blocks of causal inference -- a theme that we will further pursue in Section \ref{sec_confounding}. With the help of the d-separation criterion, their validity can be determined simply based on the topology of the graph. For this reason, DAGs constitute a valuable complement to the treatment effects literature, in which independence assumptions for counterfactuals, such as \emph{ignorability}, are usually invoked without a reference to an explicit model (\citealp{Imbens2015}). A shortcoming of such an approach is that the analyst has little to no guidance for scrutinizing the plausibility of crucial identifying assumptions on which the whole analysis hinges on. DAGs facilitate this task significantly; in particular, because finding d-separation relations, even in complex graphs, can be easily automated (\citealp{Textor2011, Textor2011a}). Moreover, using causal graphs increases the transparency of research designs compared to purely verbal justifications of identification strategies and thereby improves the communication between researchers and facilitates cumulative research efforts, as exemplified in future sections.

%%%%%%%%%%%%%%%%%%%%%%%%%%%%%
\subsection{Interventions in structural causal models}
\label{sec_interventions}
%%%%%%%%%%%%%%%%%%%%%%%%%%%%%

The aim of causal inference is to predict the effects of interventions, such as those resulting from policy actions, social programs, and management initiatives (\citealp{Woodward2003}). Based on early ideas from the econometrics literature (\citealp{Haavelmo1943,Strotz1960,Pearl2015}), interventions in structural causal models are carried out by deleting individual functions, $f_i$, from the model and fixing their left-hand side variables at a constant value.\footnote{The early literature on graphical models, including Bayesian networks and Markov random fields, relied entirely on probabilistic models, which were unable to answer causal and counterfactual queries (\citealp{Pearl2018}, p.\ 284f.). A major intellectual breakthrough was achieved in the early 1990s by switching focus to the quasi-deterministic functional relationships of the sort that are ubiquitous in econometrics (\citealp{Pearl2009}, p.\ 104f.). For a more technical discussion on the semantics and inevitability of the assumptions encoded in such models, please refer to (\citealp{bareinboim2020pch}).} As alluded to earlier, this action is denoted by a mathematical operator called $do(\cdot)$. For example, in model $M$ of equation \ref{eq2_1} (with the respective graph shown in Figure \ref{fig1a}), the action $do(X=x_0)$ results in the post-intervention model $M_{x_0}$
\begin{equation}
\label{eq2_2}
\begin{aligned}
Z &\leftarrow f_z(U_z), \\
X &\leftarrow x_0, \\
Y &\leftarrow f_y(X, Z, U_y).
\end{aligned}
\end{equation}

The diagram associated with $M_{x_0}$ is depicted in Figure \ref{fig1c}, in which all incoming arrows into $X$ are deleted and replaced by $X \leftarrow x_0$. This captures the notion that an intervention interrupts the original data generating process and eliminates all naturally occurring causes of the manipulated variable. Because other causal paths are effectively shut off in that way, any difference between two probability distributions associated with $M_{x_0}$ and $M_{x_1}$ (i.e., the system under the intervention $do(X=x_1$)) captures the variations in outcome $Y$ that is the result of a causal impact of $\Delta x = x_1 - x_0$. A randomized control trial closely follows this idea. Experimentation ties the value of a variable to the outcome of a coin flip (or randomization device), which thus induces variation in $X$ that  is uncorrelated to any other factors or causal mechanisms.

The post-intervention distribution of $Y$ can also be denoted in counterfactual notation as
\begin{equation}
\label{eq2_3}
P(y|do(x)) \triangleq P(Y_x = y), 
\end{equation}
where $Y_x = y$ should be read as ``$Y$ would be equal to $y$, if $X$ had been $x$" (\citealp{Pearl2009}, Definition 7.1.5).
This definition illustrates the connection to the potential outcomes framework (\citealp{Neyman1923,Rubin1974,Imbens2004}), where counterfactuals such as $Y_{x_0}$ and $Y_{x_1}$ are taken as primitives. By contrast, in an SCM, counterfactuals are constructs; i.e., derivable quantities from the underlying, more fundamental causal mechanisms. Naturally, we can write explicitly 
\begin{align} \label{eq2_4}
Y_{x_0} &\leftarrow f(x_0, Z, U_y), \\
Y_{x_1} &\leftarrow f(x_1, Z, U_y),
\end{align}
which immediately follow from $M_{x_0}$ and $M_{x_1}$, respectively. In other words, counterfactuals are \emph{derived} from first principles in SCMs and the corresponding causal mechanisms, instead of taken as axiomatic primitives.

Equipped with clear semantics for causal models in terms of the underlying mechanisms, and causal effects in terms of interventions on the naturally occurring structural processes in the system, we can now finally state the problem of nonparametric identification.\footnote{See \citet{Matzkin2007} and \citet{Lewbel2019} for related definitions of identifiability used in econometrics; see also footnote 10.}
\begin{definition}
	\label{def_identification}
	(Observational identifiability; \citealp{Pearl2000}) Let $Q(M)$ be any computable quantity of a model $M$. $Q$ is identifiable ($\mathpzc{ID}$, for short) from distribution $P(v)$ compatible with a causal graph $G$, if for any two (fully specified) models $M_1$ and $M_2$ that satisfy the assumptions encoded in $G$, we have
	\begin{equation}
	P_1(v) = P_2(v) \Rightarrow Q(M_1) = Q(M_2).
	\end{equation} 
\end{definition}
This definition requires that for any two (unobserved) SCMs $M_1$ and $M_2$, if their induced distributions $P_1(v)$ and $P_2(v)$ coincide, both models need to provide the same answers to query $Q$. Identifiability entails that $Q$ depends solely on $P(v)$ and the assumptions in $G$, and can therefore be uniquely expressed in terms of the observed distribution. This holds true \emph{regardless} of the underlying mechanisms $f_i$ and randomness $P(u)$, which, therefore, do not need to be known to the analyst. This is a quite remarkable result, if achieved, since while embracing and acknowledging the true, unobserved structural mechanisms, one can still make the causal statement \emph{as if} these mechanisms were fully known, such as they would be, e.g., in many settings in physics, chemistry, or biology. 

Naturally, once the post-intervention distribution $P(y|do(x))$ for any value of $x$ is identified, the average causal effect (as well as any other quantity derived from it, such as risk ratios, odds ratios, quantile effects, etc.) can be computed as\footnote{For ease of exposition, we assume random variables to be discrete throughout the text. Summations should be replaced by integrals if variables with continuous support are considered.}
\begin{equation}
E[Y|do(X = x_1)] - E[Y|do(X = x_0)] = \sum_y y  [P(y|do(x_1)) - P(y|do(x_0))].
\end{equation}

   %%%%%%%%%%%%%%%%%%%%%%%%%%%%%%%%%%%%%%%%%%%%%%
\section{Confounding Bias}
\label{sec_confounding}
%%%%%%%%%%%%%%%%%%%%%%%%%%%%%%%%%%%%%%%%%%%%%%

\setcounter{equation}{0}

One of the biggest threats to causal inference, and the one which usually receives the greatest attention from methodologists, is confounding bias. The suspicion that a correlation might not reflect a genuine causal link between two variables, but is instead driven by a set of common causes, gives rise to the maxim \emph{``correlation does not imply causation"}. In the presence of confounding, the analyst needs to find a (non-trivial) mapping from a causal query $Q$ to observables $P(v)$, in order to achieve identification. In this section, we will introduce the inference rules of \emph{do-calculus} that allow a logical and systematic treatment of the identification problem solely based on information encoded in a directed acyclic graph $G$.

Before we do so, however, we will discuss two special cases for dealing with confounding bias -- backdoor and frontdoor adjustment -- that are instances of the general treatment provided by do-calculus. Eventually, we will also discuss identification strategies for cases when confounding bias cannot be eliminated in purely observational data, but in which a surrogate experiment (akin to an instrumental variable that creates exogenous variation in a treatment) is available. 

%%%%%%%%%%%%%%%%%%%%%%%%%%%%%
\subsection{Covariate selection and the backdoor criterion}
\label{sec_backdoor}
%%%%%%%%%%%%%%%%%%%%%%%%%%%%%

\setcounter{equation}{0}

\begin{figure}[thp]
	\centering
	\subcaptionbox{\label{fig_college}}[.9\linewidth]{
		\begin{tikzpicture}[>=triangle 45, font=\footnotesize]
		% Nodes
		\node[fill,circle,inner sep=0pt,minimum size=5pt,label={left:{C}}] (C) at (0,0) {};
		\node[fill,circle,inner sep=0pt,minimum size=5pt,label={right:{Y}}] (Y) at (4,0) {};
		\node[fill,circle,inner sep=0pt,minimum size=5pt,label={above:{W}}] (W) at (2,2) {};
		\node[fill,circle,inner sep=0pt,minimum size=5pt,label={[xshift=-0.2cm, yshift=0.1cm]below:{H}}] (H) at (2,0.6) {};
		\node[fill,circle,inner sep=0pt,minimum size=5pt,label={below:{E}}] (E) at (2,-1) {};
		% Arcs
		\draw[->,shorten >= 1pt] (C)--(Y);
		\draw[->,shorten >= 1pt,] (C)--(W);
		\draw[->,shorten >= 1pt] (W)--(H);
		\draw[->,shorten >= 1pt] (W)--(Y);
		\draw[->,shorten >= 1pt] (H)--(Y);
		\draw[->,shorten >= 1pt] (E)--(C);
		\draw[->,shorten >= 1pt] (E)--(Y);
		\draw[<->,dashed,shorten >= 1pt] (W) to[bend left=45] (Y);
		\draw[<->,dashed,shorten >= 1pt] (C) to[bend right=45] (E);
		\end{tikzpicture}
	} \\
	\subcaptionbox{\label{fig_mutilated1}}[.45\linewidth]{
		\begin{tikzpicture}[>=triangle 45, font=\scriptsize, scale=0.8]
		\node[fill,circle,inner sep=0pt,minimum size=5pt,label={left:{C}}] (C) at (0,0) {};
		\node[fill,circle,inner sep=0pt,minimum size=5pt,label={right:{Y}}] (Y) at (4,0) {};
		\node[fill,circle,inner sep=0pt,minimum size=5pt,label={above:{W}}] (W) at (2,2) {};
		\node[fill,circle,inner sep=0pt,minimum size=5pt,label={[xshift=-0.2cm, yshift=0.1cm]below:{H}}] (H) at (2,0.6) {};
		\node[fill,circle,inner sep=0pt,minimum size=5pt,label={below:{E}}] (E) at (2,-1) {};
		\draw[->,shorten >= 1pt] (W)--(H);
		\draw[->,shorten >= 1pt] (W)--(Y);
		\draw[->,shorten >= 1pt] (H)--(Y);
		\draw[->,shorten >= 1pt] (E)--(C);
		\draw[->,shorten >= 1pt] (E)--(Y);
		\draw[<->,dashed,shorten >= 1pt] (W) to[bend left=45] (Y);
		\draw[<->,dashed,shorten >= 1pt] (C) to[bend right=45] (E);
		\end{tikzpicture}
	}
	\subcaptionbox{\label{fig_mutilated2}}[.45\linewidth]{
		\begin{tikzpicture}[>=triangle 45, font=\scriptsize, scale=0.8]
		\node[fill,circle,inner sep=0pt,minimum size=5pt,label={left:{C}}] (C) at (0,0) {};
		\node[fill,circle,inner sep=0pt,minimum size=5pt,label={right:{Y}}] (Y) at (4,0) {};
		\node[fill,circle,inner sep=0pt,minimum size=5pt,label={above:{W}}] (W) at (2,2) {};
		\node[fill,circle,inner sep=0pt,minimum size=5pt,label={[xshift=-0.2cm, yshift=0.1cm]below:{H}}] (H) at (2,0.6) {};
		\node[fill,circle,inner sep=0pt,minimum size=5pt,label={below:{E}}] (E) at (2,-1) {};
		\draw[->,shorten >= 1pt] (C)--(Y);
		\draw[->,shorten >= 1pt,] (C)--(W);
		\draw[->,shorten >= 1pt] (W)--(H);
		\draw[->,shorten >= 1pt] (W)--(Y);
		\draw[->,shorten >= 1pt] (H)--(Y);
		\draw[->,shorten >= 1pt] (E)--(Y);
		\draw[<->,dashed,shorten >= 1pt] (W) to[bend left=45] (Y);
		\end{tikzpicture}
	}
	\caption{A graphical causal model $G$ for estimating the college wage premium (together with the mutilated graphs $G_{\underline{C}}$ (b) and $G_{\overline{C}}$ (c) used in the do-calculus derivations in Section \ref{sec_do-calculus}).}
\end{figure}

Consider the well-known example from labour economics of estimating the college wage premium (\citealp{Angrist2009}, ch. 3.2.3). Let the causal relationships in the problem be represented by the causal graph $G$ in Figure \ref{fig_college}. $C$ is a dummy variable that is equal to one for individuals who obtained a college degree, and the outcome of interest, $Y$, refers to annual earnings. $W$ is a dummy indicating whether an individual works in a ``white-collar" or ``blue-collar" job. $W$ is causally affected by $C$, since many white-collar jobs require a college degree. At the same time, the effect of $W$ is partially mediated by an individual's work-related health $H$. This assumption captures the idea that blue-collar jobs might be associated with relatively higher adverse health effects, which ultimately reduce life-time earnings. Finally, $E$ represents a set of socioeconomic variables that influence both the probability to graduate from college as well as individuals' future earning potentials. Dashed bidirected arrows depict unmeasured common causes that lead to a dependence between the background factors $U$ associated with the connected variables.

In order to estimate the causal effect of a college degree on earnings, the following graphical criterion can be used to find admissible adjustment sets that eliminate any confounding influences between $C$ and $Y$.

\begin{definition}\label{def_backdoor}
	(Admissible sets -- the backdoor criterion; \citealp{Pearl1995})
	Given an ordered pair of treatment and outcome variables $(X,Y)$ in a directed acyclic graph $G$, a set $Z$ is backdoor-admissible if it blocks every path between $X$ and $Y$ in the graph $G_{\underline{X}}$.
\end{definition}

$G_{\underline{X}}$ in Definition \ref{def_backdoor} refers to the graph that is obtained when all edges emitted by node $X$ are deleted in $G$. Figure \ref{fig_mutilated1} depicts the modified graph $G_{\underline{C}}$ for the college wage premium example, where the edges $C \rightarrow Y$ and $C \rightarrow W$ have been removed. The intuition behind the backdoor criterion is simple. Unblocked paths between $X$ and $Y$ pointing into $X$ (i.e., they ``enter through the backdoor") create an association between $X$ and $Y$ that is not due to any causal influence exerted by $X$.\footnote{Genuine causal effects can only be transmitted ``downstream" of $X$, via directed paths pointing from $X$ to its descendants and  eventually to $Y$.} By adjusting for variables along these paths, this spurious association can be eliminated such that only the causal effect of $X$ on $Y$ remains. 

In the particular example of Figure \ref{fig_college}, the set $Z=\{E\}$ satisfies the backdoor criterion and is thus an admissible adjustment set.\footnote{Note that $Z=\{E\}$ remains an admissible adjustment set even if edges pointing from $E$ to $W$ and $H$ are added to the graph in Figure \ref{fig_college}.} $W$ can be left unaccounted for because it does not lie on a backdoor path between $X$ and $Y$. In fact, the graph illustrates why conditioning on occupation would produce, rather than reduce, estimation bias. According to the d-separation criterion in Definition \ref{def_d-separation}, $W$ is a collider node on $C \rightarrow W \dashleftarrow \dashrightarrow Y$, and thus would open, or unblock, this path when conditioned on. As a consequence, adjusting for $W$ would inject bias in this example, creating a non-causal (spurious) correlation between $C$ and $Y$, and would therefore be a serious mistake.

Whenever a backdoor-admissible set exists, the causal effect of $X$ on $Y$ can, in principle, be estimated by adjustment, as shown next. 

\begin{theorem} (Backdoor adjustment criterion) 
\label{th_backdoor_adjustment_formula}
If a set of variables $Z$ satisfies the backdoor criterion relative to $(X, Y)$, the causal effect of $X$ on $Y$ can be identified from observational data by the adjustment formula 
\begin{equation}\label{adjustment_formula}
P(Y=y|do(X=x)) = \sum_{z} P(Y=y|X=x,Z=z)P(Z=z).
\end{equation}
\end{theorem}
\noindent Practically speaking, estimation can be carried out by propensity score matching (\citealp{Rosenbaum1983,Heckman1998}), inverse probability weighting (\citealp{Horvitz1952,Robins1999}), deep neural networks (\citealp{shi:etal19}),  or weighted empirical risk minimization (\citealp{Jung2020}), among other efficient estimation methods. 
If the cardinality of $Z$ is high, regularization techniques such as the double machine learning framework by \citet{Chernozhukov2018} can be applied, including for arbitrary graph structures (\citealp{jung2021dmlid}).

At this point, the similarity with the treatment effects literature is no coincidence, as the backdoor criterion formally implies \emph{ignorability} (\citealp{Rosenbaum1983}), as shown next. 

\begin{theorem} 
	\label{th_counterfactual_backdoor}
	(Counterfactual interpretation of backdoor; \citealp{Pearl2009}) If a set of variables $Z$ satisfies the backdoor criterion relative to $(X,Y)$, then for all $x$, the counterfactual $Y_x$ is conditionally independent of $X$ given $Z$:
	\begin{equation}
	Y_x \independent X | Z.
	\end{equation}
\end{theorem}

\begin{figure}[thp]
	\centering
	\begin{tikzpicture}[>=triangle 45, font=\footnotesize]
		% Nodes
		\node[fill,circle,inner sep=0pt,minimum size=5pt,label={left:{RDSUB}}] (RDSUB) at (0,0) {};
		\node[fill,circle,inner sep=0pt,minimum size=5pt,label={right:{RDINT}}] (RDINT) at (8,0) {};
        \node[fill,circle,inner sep=0pt,minimum size=5pt,label={below:{AC}}] (AC) at (4,0) {};
        \node[fill,circle,inner sep=0pt,minimum size=5pt,label={[xshift=-0.1cm, yshift=-0.05cm]below:{FCONS}}] (FCONS) at (4,1) {};
        \node[fill,circle,inner sep=0pt,minimum size=5pt,label={above:{PQUAL}}] (PQUAL) at (4,2.2) {};
        \node[fill,circle,inner sep=0pt,minimum size=5pt,label={left:{EMP}}] (EMP) at (1,3) {};
        \node[fill,circle,inner sep=0pt,minimum size=5pt,label={left:{AGE}}] (AGE) at (2.5,3) {};
        \node[fill,circle,inner sep=0pt,minimum size=5pt,label={above:{EXP}}] (EXP) at (5,4) {};
		\node[fill,circle,inner sep=0pt,minimum size=5pt,label={right:{IND}}] (IND) at (7,3) {};
		% Arcs
		\draw[->,shorten >= 1pt] (RDSUB)--(AC);
        \draw[->,shorten >= 1pt] (RDINT)--(AC);
        \draw[->,shorten >= 1pt] (FCONS)--(RDINT);
        \draw[->,shorten >= 1pt] (PQUAL)--(RDINT);
        \draw[->,shorten >= 1pt] (PQUAL)--(RDSUB);
        \draw[->,shorten >= 1pt] (EMP)--(RDSUB);
        \draw[->,shorten >= 1pt] (EMP)--(FCONS);
        \draw[->,shorten >= 1pt] (EMP) to[bend left=20] (EXP);
        \draw[->,shorten >= 1pt] (AGE)--(FCONS);
        \draw[->,shorten >= 1pt] (AGE)--(EXP);
        \draw[->,shorten >= 1pt] (AGE)--(RDSUB);
        \draw[->,shorten >= 1pt] (EXP)--(RDINT);
        \draw[->,shorten >= 1pt] (IND)--(AGE);
        \draw[->,shorten >= 1pt] (IND)--(EXP);
        \draw[->,shorten >= 1pt] (IND)--(FCONS);
        \draw[->,shorten >= 1pt] (IND)--(PQUAL);
		\draw[->,shorten >= 1pt] (RDSUB) to[bend right=30] (RDINT);
	\end{tikzpicture}
	\caption{A graphical model of R\&D subsidy additionality. $AC$: absorptive capacity, $AGE$: firm age, $EMP$: number of employees, $EXP$: exporter, $FCONS$: financial constraints, $IND$: industry, $PQUAL$: project quality, $RDINT$: R\&D intensity, $RDSUB$: R\&D subsidy.}
\label{fig_large_graph}
\end{figure}

In contrast to the potential outcomes framework, however, which provides the analyst with little guidance on identifying biasing paths and admissible sets (Definition \ref{def_backdoor}), the search for appropriate adjustment sets via the backdoor criterion can easily be automated (\citealp{Textor2011, Textor2011a}). This is particularly useful in larger graphs such as in Figure \ref{fig_large_graph}, which presents a model of input additionality of R\&D subsidies. The model stipulates that research grants ($RDSUB$) are assigned based on the quality of projects with which firms apply ($PQUAL$). In addition, young ($AGE$) as well as small and medium-sized enterprises (defined as having fewer than 250 employees in the EU, $EMP$) often receive preferential treatment under many policy regimes (\citealp{Howell2017,Huenermund2019}). Other important covariates in this context are whether firms are exporters ($EXP$, \citealp{Peters2018}), the degree to which they are financially constrained ($FCONS$, \citealp{Hottenrott2012}), their level of absorptive capacity ($AC$, \citealp{Cohen1989}), and the industry they are operating in ($IND$). The outcome of interest is whether R\&D subsidies are able to raise firms' R\&D intensity ($RDINT$), measured as R\&D expenditures in percentage of sales.

The two smallest (minimum) admissible adjustment sets for identifying the effect of $RDSUB$ on $RDINT$ in Figure \ref{fig_large_graph} are given by 
\begin{equation}
    \{EXP, FCONS, PQUAL\} \quad \text{and} \quad \{AGE, EMP, PQUAL\}.
\end{equation}
This example illustrates that it is neither necessary nor sufficient to adjust for all covariates in a model. The analyst could, for example, decide that financial constraints at the firm-level are too difficult to measure and instead proceed with the second admissible adjustment set.

At the same time, it would be a mistake to condition on the node $AC$. Absorptive capacity is the result of firms' R\&D investments (\citealp{Cohen1990}, p.\ 141). Furthermore, there is evidence that R\&D subsidies are able to increase absorptive capacity through a process that is commonly referred to as behavioral additionality (\citealp{Clarysse2009}). Taken together, these two assumptions turn $AC$ into a collider, which, according to the d-separation criterion (Definition \ \ref{def_d-separation}), would open up the path between $RDSUB$ and $RDINT$ and lead to bias if conditioned on. The largest admissible adjustment set is thus given by $\{AGE, EMP, EXP, FCONS, IND, PQUAL\}$, which could be preferred over a smaller set for reasons of estimation efficiency (\citealp{White2011,Cinelli2022}). These intricacies are hard to spot without the use of a causal diagram. We therefore regard DAGs as a useful complement to reduced-form approaches that mainly rely on verbal theorizing. More formalized approaches can likewise benefit from the visualization and automation properties of causal diagrams, which we will further describe below.

%%%%%%%%%%%%%%%%%%%%%%%%%%%%%
\subsection{Frontdoor adjustment in the presence of unmeasured confounders}
\label{sec_frontdoor}
%%%%%%%%%%%%%%%%%%%%%%%%%%%%%

\begin{figure}[thp]
	\centering
	\subcaptionbox{\label{fig_frontdoor1}}[.48\linewidth]{
		\begin{tikzpicture}[>=triangle 45, font=\footnotesize]
		\node[fill,circle,inner sep=0pt,minimum size=5pt,label={below:{X}}] (X) at (0,1) {};
		\node[fill,circle,inner sep=0pt,minimum size=5pt,label={below:{Y}}] (Y) at (4,1) {};
		\node[fill,circle,inner sep=0pt,minimum size=5pt,label={below:{M}}] (M) at (2,1) {};
		\node[] (placeholder) at (0,0) {};
		\draw[->,shorten >= 1pt] (X)--(M);
		\draw[->,shorten >= 1pt] (M)--(Y);
		\draw[<->,dashed,shorten >= 1pt] (X) to[bend left=60] (Y);
		\end{tikzpicture}
	}
	\subcaptionbox{\label{fig_frontdoor2}}[.48\linewidth]{
		\begin{tikzpicture}[>=triangle 45, font=\footnotesize]
		\node[fill,circle,inner sep=0pt,minimum size=5pt,label={left:{X}}] (X) at (0,0) {};
		\node[fill,circle,inner sep=0pt,minimum size=5pt,label={right:{Y}}] (Y) at (4,0) {};
		\node[fill,circle,inner sep=0pt,minimum size=5pt,label={below:{M}}] (M) at (2,0) {};
		\node[fill,circle,inner sep=0pt,minimum size=5pt,label={above:{W$_2$}}] (W2) at (1,1) {};
		\node[fill,circle,inner sep=0pt,minimum size=5pt,label={above:{W$_1$}}] (W1) at (2,2) {};
		\node[fill,circle,inner sep=0pt,minimum size=5pt,label={above:{W$_3$}}] (W3) at (3,1) {};
		\draw[->,shorten >= 1pt] (X)--(M);
		\draw[->,shorten >= 1pt] (M)--(Y);
		\draw[->,shorten >= 1pt] (W2)--(X);
		\draw[->,shorten >= 1pt] (W2)--(M);
		\draw[->,shorten >= 1pt] (W1) to[bend right=45] (X);
		\draw[->,shorten >= 1pt] (W1) to[bend left=45] (Y);
		\draw[->,shorten >= 1pt] (W3) -- (M);
		\draw[->,shorten >= 1pt] (W3) -- (Y);
		\draw[<->,dashed,shorten >= 1pt] (X) to[bend right=45] (Y);
		\end{tikzpicture}
	}
	\caption{The frontdoor criterion.}
\label{fig_frontdoor}
\end{figure}

Identification via backdoor adjustment requires that all backdoor paths can be blocked by a set of observed nodes, which might not always be feasible. In situations where no set of observables is backdoor-admissible, another (somewhat less familiar to economists) identification strategy might be applicable. Figure \ref{fig_frontdoor1} presents an example in which adjusting for a set of observable variables is not sufficient to close all backdoor paths between $X$ and $Y$. For any possible adjustment set, there are unobserved confounders remaining in the graph, represented by the bidirected arc $X \dashleftarrow \dashrightarrow Y$. At the same time, the entire effect of $X$ is assumed to be mediated by another observed variable $M$ and there are no unobserved confounders connecting $M$ with $X$ and $Y$. These assumptions may be plausible in settings in which a test result $M$ provides a noisy signal for the presence of a characteristic $X$. For example, congenital anomalies are routinely tested via ultrasound screenings during pregnancy. However, these screenings exhibit a positive type-1 and type-2 error rate (\citealp{Debost-Legrand2014}). If diagnostic missclassifications occur randomly, or are related to observables such as maternal BMI, confounding at the mediator can be controlled for. Full mediation in this case implies that congenital anomalies will only affect the likelihood of an outcome $Y$, such as the decision to terminate the pregnancy, if they are detected during an ultrasound screening.\footnote{In case other prenatal testing methods than ultrasound are applied that could lead to the detection of congenital anomalies, they should be included as additional mediators in the model. By contrast, if congenital anomalies affect other unobservable markers that lead to pregnancy termination, the assumption of full mediation would be violated due to a direct (from the perspective of the analyst) effect $X \rightarrow Y$.}

In this setting, the causal effect $P(y|do(x))$ is identifiable with the help of the following criterion (generalizing Theorem 2 in \citealp{Pearl1995}).
 
\begin{definition}\label{def_frontdoor}(Conditional frontdoor criterion) A set of variables $M$ is said to satisfy the conditional frontdoor criterion (frontdoor, for short) relative to a triplet $(X,Y, W)$ if: (a) $M$ intercepts all directed paths from $X$ to $Y$, (b) there is no unblocked backdoor path from $X$ to $M$ given $W$, and (c) all the backdoor paths from $M$ to $Y$ are blocked by $\{X, W\}$.
\end{definition}

\begin{theorem}(Conditional frontdoor adjustment) If a set of variables $M$ satisfies the conditional frontdoor criterion relative to $(X, Y, W)$, the causal effect of $X$ on $Y$ can be identified from observational data by the frontdoor formula
\begin{equation}\label{frontdoor_adjustment_formula}
P(Y=y|do(X=x)) = \sum_{m, w} P(m | w, X=x) p(w) \sum_{x'}  P(Y=y|w, m, X=x') P(X= x' | w).
\end{equation}
\end{theorem}

Frontdoor adjustment amounts to a sequential application of the backdoor criterion. In Figure \ref{fig_frontdoor2}, first, the effect of $X$ on $M$ can be identified by adjusting for $W_2$. Second, the backdoor path $M \leftarrow X \dashleftarrow \dashrightarrow Y$, which remains open after adjusting for $W_1$ and $W_3$, can be blocked by conditioning on $X$, to identify the effect of $M$ on $Y$. The frontdoor adjustment formula then chains these individual causal effect estimates together to arrive at the overall effect of $X$ on $Y$. Because the frontdoor criterion is applicable in the presence of direct unobserved confounding between treatment and outcome (i.e., ignorability does not hold), it is a good example of how causal graphs can point to new identification strategies that go beyond the standard tools currently applied in econometrics.\footnote{\citet{Glynn2017} present an interesting application of the frontdoor criterion (FDC) for evaluating the effect of the National Job Training Partnership Act program (\citealp{Heckman1997}) on earnings by complementing the FDC with a difference-in-differences-type identification approach that tackles potential bias stemming from unobserved confounders between $M$ and $Y$. \citet{Bellemare2020} apply the FDC to study how ride sharing affects tipping behavior on popular ride-hailing apps such as Uber and Lyft.}

%%%%%%%%%%%%%%%%%%%%%%%%%%%%%                    
\subsection{Causal calculus and the algorithmatization of identification strategies}                                                                          
\label{sec_do-calculus}                                                                                                                          
%%%%%%%%%%%%%%%%%%%%%%%%%%%%%                                                                                                                    
The backdoor and frontdoor criteria offer simple graphical identification rules that are easy to check in the causal diagram. However, while definitely important, they only represent a very limited subset of the overall identification results that are derivable in DAGs. In more generality, identifiability of any query of the form $P(y|do(x))$ can be decided systematically by using a symbolic causal inference engine called \emph{do-calculus} (\citealp{Pearl1995}). Do-calculus consists of three inference rules that allow the analyst to transform probabilistic sentences involving interventions and observations, whenever certain separation conditions hold in the causal graph $G$ defined by model $M$. 

Let $X$, $Y$, $Z$, and $W$ be arbitrary disjoint sets of nodes in $G$. The mutilated graph that is obtained by removing all arrows pointing to nodes in $X$ from $G$ is denoted by $G_{\overline{X}}$. Similarly, $G_{\underline{X}}$ results from deleting all arrows that are emitted by $X$ in $G$. Finally, the removal of both arrows incoming in $X$ and arrows outgoing from $Z$ is denoted by $G_{\overline{X}\underline{Z}}$. Given this notation, the following three rules -- valid for every interventional distribution compatible with $G$ -- can be formulated.
        
\begin{rules}                                                          
	(Insertion/deletion of observations)                                                                                                            
		\begin{equation}                                                        
		P(y | do(x), z, w) = P(y | do(x), w) \qquad \text{\normalfont if} \enskip (Y \independent Z | X, W)_{G_{\overline{X}}}.
		\end{equation}                                                                                                                                  
\end{rules} \vspace{-\baselineskip}

\begin{rules}
	(Action/observation exchange)
	\begin{equation}
	P(y | do(x), do(z), w) = P(y | do(x), z, w) \qquad \text{\normalfont if} \enskip (Y \independent Z | X, W)_{G_{\overline{X}\underline{Z}}}.
	\end{equation}
\end{rules} \vspace{-\baselineskip}

\begin{rules}
	(Insertion/deletion of actions)
	\begin{equation}
	P(y | do(x), do(z), w) = P(y | do(x), w) \qquad \text{\normalfont if} \enskip (Y \independent Z | X, W)_{G_{\overline{XZ(W)}}},
	\end{equation}
	where $Z(W)$ is the set of Z-nodes that are not ancestors of any W-node in $G_{\overline{X}}$.
\end{rules}

\smallskip 
\noindent 

Rule 1 is a reaffirmation of the d-separation criterion for the $X$-manipulated graph $G_{\overline{X}}$, i.e., for any interventional distribution $do(X)$. Since $Z$ is independent of $Y$, conditional on $X$ and $W$, $Z$ can be freely inserted or deleted in the do-expression. Rule 2 states the condition for an intervention $do(Z=z)$ to have the same effect as a passively observed $Z=z$. This condition is fulfilled if $\{X \cup W\}$ blocks all backdoor paths from $Z$ to $Y$. Note that in $G_{\overline{X}\underline{Z}}$ only such backdoor paths are remaining, since edges emitted by $Z$ are deleted from the graph. Rule 3, then indicates under which conditions a manipulation of $Z$ does not affect the probability of $Y$. Roughly speaking, this is the case if in the $X$- and $Z$-manipulated graph $G_{\overline{XZ}}$, $Z$ is independent of $Y$ conditional on $X$ and $W$.\footnote{The reason for restricting the deletion to $Z$-nodes that are not ancestors of any $W$-node in rule 3 of the do-calculus is provided with the proofs in \citet{Pearl1995}.}

Identifiability of a causal query can be decided by repeatedly applying the rules of do-calculus, until $Q$ is transformed into a final expression that no longer contains a do-operator. This provides the basis for consistent estimation of $Q$ from nonexperimental data. In the following, we demonstrate this process by showing a step-by-step do-calculus derivation for the causal effect of $C$ on $Y$ in the college wage premium example in Figure \ref{fig_college}. Notice that there are two backdoor paths in Figure \ref{fig_college}, which can both be blocked by $E$. By the law of total probability, $P(y|do(c))$ can be written as
\begin{equation}
P(y|do(c)) = \sum_{e}P(y|do(c),e)P(e|do(c)).
\end{equation}
By rule 2 of do-calculus, since $(Y \independent C | E)$ in subgraph $G_{\underline{C}}$, it holds that
\begin{equation}
P(y|do(c),e) = P(y|c,e).
\end{equation}
In $G_{\overline{C}}$, $E$ is d-separated from $C$, because $Y$ is a collider on every path connecting them. Thus, $(E \independent C)_{G_{\overline{C}}},$ and by rule 3 of do-calculus
\begin{equation}
P(e|do(c)) = P(e).
\end{equation}
Combining these two expressions yields
\begin{equation*}
P(y|do(c)) = \sum_{e}P(y|c,e)P(e).
\end{equation*}
The right-hand-side expression is do-free and can therefore -- in principle -- be estimated from nonexperimental data.

Do-calculus was proved sound and complete for general queries of the form $Q = P(y|do(x),z)$ (\citealp{Pearl1995,tian2002general,Shpitser2006,Huang2006,Bareinboim2012,Lee:etal19}) from a combination of observational and experimental data. Soundness assures that an answer returned by do-calculus is correct. Completeness means that do-calculus is guaranteed to return a solution for the identification problem, whenever such a solution exists. It implies that if no sequence of steps applying the rules of do-calculus can be found that allow to transform $Q$ into an expression only consisting of ex-post observed probabilities, the causal effect is known to be non-identifiable with observational data. If that is the case, point identification will only be achievable by imposing stronger functional form restrictions (such as linearity, monotonicity, additivity, etc.) or by making assumptions about the distribution of the background factors $U$. In fact, this result can also be seen algorithmically, which allows one to fully automate the often tedious task of transforming causal effect queries into do-free expressions. This way, the identification of causal effects becomes a straightforward exercise that can be solved with the help of a computer (\citealp{Tian2002}).\footnote{More recently, for effects that are provably not point identifiable,  very general machinery for partial identification has been  developed that  can be applied from any causal diagram and any arbitrary combination of observational and experimental distributions (\citealp{Zhang2022}).}

%%%%%%%%%%%%%%%%%%%%%%%%%%%%%
\subsection{Identification by surrogate experiments}
\label{sec_z-identifiability}
%%%%%%%%%%%%%%%%%%%%%%%%%%%%%

\begin{figure}[thp]
	\centering
	\subcaptionbox{\label{fig_zid1}}[.45\linewidth]{
	    \begin{tikzpicture}[>=triangle 45, font=\footnotesize]
		\node[fill,circle,inner sep=0pt,minimum size=5pt,label={below:{X}}] (X) at (0,1) {};
		\node[fill,circle,inner sep=0pt,minimum size=5pt,label={below:{Y}}] (Y) at (3.4,1) {};
		\node[fill,circle,inner sep=0pt,minimum size=5pt,label={[xshift=0.05cm, yshift=0.05cm]above:{Z}}] (Z) at (1.7,2.5) {};
		\node[] (placeholder) at (0,0) {};
		\draw[->,shorten >= 1pt] (Z)--(X);
		\draw[->,shorten >= 1pt] (X)--(Y);
		\draw[<->,dashed,shorten >= 1pt] (Z) to[bend left=60] (Y);
		\draw[<->,dashed,shorten >= 1pt] (Z) to[bend right=75] (X);
		\end{tikzpicture}
	}
	\subcaptionbox{\label{fig_zid2}}[.45\linewidth]{
		\begin{tikzpicture}[>=triangle 45, font=\footnotesize]
		% Nodes
		\node[fill,circle,inner sep=0pt,minimum size=5pt,label={above:{W$_1$}}] (W1) at (0,3) {};
		\node[fill,circle,inner sep=0pt,minimum size=5pt,label={above:{Z}}] (Z) at (1.5,3) {};
		\node[fill,circle,inner sep=0pt,minimum size=5pt,label={right:{X}}] (X) at (2.1,1.8) {};
		\node[fill,circle,inner sep=0pt,minimum size=5pt,label={below:{Y}}] (Y) at (3,0) {};
		\node[fill,circle,inner sep=0pt,minimum size=5pt,label={[xshift=0.14cm, yshift=0.25cm]left:{W$_2$}}] (W2) at (1.7,0.6) {};
		% Arcs
		\draw[->,shorten >= 1pt] (W1)--(Z);
		\draw[->,shorten >= 1pt] (Z)--(X);
		\draw[->,shorten >= 1pt] (X)--(Y);
		\draw[->,shorten >= 1pt] (W2)--(X);
		\draw[->,shorten >= 1pt] (W2)--(Y);
		\draw[->,shorten >= 1pt] (W1) to[out=270, in=190] (Y);
		\draw[<->,dashed,shorten >= 1pt] (Z) to[bend left=60] (X);
		\draw[<->,dashed,shorten >= 1pt] (Z) to[bend left=60] (Y);
		\draw[<->,dashed,shorten >= 1pt] (W1) to[bend right=20] (X);
		\end{tikzpicture}
	} \\
	\subcaptionbox{\label{fig_zid3}}[.45\linewidth]{
		\begin{tikzpicture}[>=triangle 45, font=\footnotesize]
		% Nodes
		\node[fill,circle,inner sep=0pt,minimum size=5pt,label={above:{W$_1$}}] (W1) at (0,3) {};
		\node[fill,circle,inner sep=0pt,minimum size=5pt,label={above:{Z}}] (Z) at (1.5,3) {};
		\node[fill,circle,inner sep=0pt,minimum size=5pt,label={right:{X}}] (X) at (2.1,1.8) {};
		\node[fill,circle,inner sep=0pt,minimum size=5pt,label={below:{Y}}] (Y) at (3,0) {};
		\node[fill,circle,inner sep=0pt,minimum size=5pt,label={[xshift=0.14cm, yshift=0.25cm]left:{W$_2$}}] (W2) at (1.7,0.6) {};
		% Arcs
		\draw[->,shorten >= 1pt] (Z)--(X);
		\draw[->,shorten >= 1pt] (X)--(Y);
		\draw[->,shorten >= 1pt] (W2)--(X);
		\draw[->,shorten >= 1pt] (W2)--(Y);
		\draw[->,shorten >= 1pt] (W1) to[out=270, in=190] (Y);
		\draw[<->,dashed,shorten >= 1pt] (W1) to[bend right=20] (X);
		\end{tikzpicture}
	}
	\subcaptionbox{\label{fig_zid4}}[.45\linewidth]{
		\begin{tikzpicture}[>=triangle 45, font=\footnotesize]
		\node[fill,circle,inner sep=0pt,minimum size=5pt,label={below:{X}}] (X) at (1,1) {};
		\node[fill,circle,inner sep=0pt,minimum size=5pt,label={below:{Y}}] (Y) at (4,1) {};
		\node[fill,circle,inner sep=0pt,minimum size=5pt,label={above:{Z}}] (Z) at (0,2.5) {};
		\node[] (placeholder) at (0,0) {};
		\draw[->,shorten >= 1pt] (Z)--(X);
		\draw[->,shorten >= 1pt] (X)--(Y);
		\draw[<->,dashed,shorten >= 1pt] (X) to[bend left=60] (Y);
		\end{tikzpicture}
	}
	\caption{Identification problems involving surrogate experiments.}
	\label{fig4}
\end{figure}

In practice, identification of causal queries based on observational data alone often remains an unattainable goal. At the same time, conducting a randomized control trial (RCT) for the treatment of interest might likewise be infeasible due to cost, ethical, or technical considerations. In such cases, a frequently applied strategy is to make use of experiments involving a third variable, which is only proximately linked to the treatment but more easily manipulable. Such surrogate experiments are sometimes referred to as \emph{``encouragement designs"} in economics (\citealp{Duflo2008}). %\citet{Duflo2003}, for example, study the effect of financial knowledge on retirement planning decisions by randomly allocating monetary rewards for attending an information session on tax deferred account (TDA) retirement plans to university employees. In this surrogate experiment, experimental control of a proxy variable (financial rewards) is supposed to create (or ``encourage") exogenous variation in the otherwise endogenous treatment of interest (knowledge about TDA retirement plans). 

Consider the situation in Figure \ref{fig_zid1}, in which $X$ represents participation in a financial support program that allows job seekers to reimburse private expenses incurred for travel and materials during the application process. $Y$ measures the employment status of an individual six months after the finish of the program. Finally, $Z$ is an indicator for whether an individual is aware of the existence of the program. For simplicity, we assume that the financial incentives provided by the program are sufficiently attractive, such that awareness is the only parent node of $X$ in the diagram (but the relationship is not deterministic). However, awareness is itself influenced by unobservables that create an unblocked backdoor path $X \leftarrow Z \dashleftarrow\dashrightarrow Y$. Unfortunately, adjusting for $Z$ is not sufficient, as this would open up the path $X \dashleftarrow\dashrightarrow Z \dashleftarrow\dashrightarrow Y$ on which $Z$ is a collider. Thus, $P(y|do(x))$ is not identifiable via backdoor adjustment in this example. Program participation cannot be forced, which rules out a direct manipulation of $X$. Nonetheless, if the analyst is able to manipulate awareness $Z$ instead, e.g., by sending information about the program only to a randomly selected group of individuals, identification becomes possible.\footnote{Perfect manipulation of $Z$ implies that individuals cannot learn about the existence of the program via different channels and that there is no communication between individuals, which may or may not be plausible given the context.} We will illustrate such a strategy involving auxiliary experiments on ancestor nodes of a treatment $X$ in the following.

To make the theory of surrogate experiments in causal diagrams more concrete, Figure \ref{fig_zid2} presents an example in which several paths passing through $Z$ are confounding the relationship between $X$ and $Y$. Backdoor adjustment is not a viable identification strategy in this graph, since $Z$ is a collider on $X \dashleftarrow\dashrightarrow Z \dashleftarrow\dashrightarrow Y$, and conditioning on $Z$ would thus open up the path. Furthermore, it can be shown that any other attempt of identifying $Q = P(y|do(x))$ with purely observational data is prone to fail as well in this example. By contrast, if it is possible to manipulate $Z$ in a randomized control trial, the causal effect of $X$ on $Y$ can be identified from the interventional distribution $P(v|do(z))$ instead. Generalizing this idea leads to a natural refinement of the identification problem formulated earlier (see Definition \ref{def_identification}).

\begin{definition}
	($\mathpzc{Z}$-identifiability; \citealp{Bareinboim2012})
	Let $X, Y, Z$ be disjoint sets of variables, and let $G$ be the causal diagram. The causal effect of an action $do(X=x)$ on a set of variables $Y$ is said to be $\mathpzc{z}$-identifiable ($\mathpzc{zID}$, for short) from $P$ in $G$, if $P(y|do(x))$ is (uniquely) computable from $P(V)$ together with the interventional distributions $P(V \setminus Z'|do(Z'))$, for all $Z' \subseteq Z$, in any model that induces $G$.
\end{definition}

\citet{Bareinboim2012} show that the $\mathpzc{z}$-identification task can be solved in a similar fashion to the standard identification problem, by repeatedly applying the rules of do-calculus in order to transform a causal query $Q$ into an expression that only contains $do(z)$.

\begin{theorem}
	\label{th_zid1}
	(\citealp{Bareinboim2012}) Let $X, Y, Z$ be disjoint sets of variables, and let $G$ be the causal diagram, and $Q = P(y|do(x))$. $Q$ is $\mathpzc{zID}$ from $P$ in $G$ if the expression $P(y|do(x))$ is reducible, using the rules of do-calculus, to an expression in which only elements of $Z$ may appear as interventional variables. 
\end{theorem}
\noindent It can further be proved that do-calculus is likewise complete for $\mathpzc{z}$-identification (\citealp{Bareinboim2012}, Corrolary 3; \citealp{Lee:etal19}, Theorem 3); i.e., it reaches a solution to the $\mathpzc{zID}$ problem whenever such a solution exists.

For the sake of concreteness, however, we discuss a weaker condition, which is only sufficient but not necessary, in order to exemplify the mechanics of the  $\mathpzc{z}$-identification problem. 

\begin{theorem}
	\label{th_zid2}
	(Sufficient condition --  $\mathpzc{z}$-identification; \citealp{Bareinboim2012})
	Let $X$, $Y$, $Z$ be disjoint sets of variables and let $G$ be the causal graph. The causal effect $Q = P(y|do(x))$ is $\mathpzc{zID}$ in $G$ if one of the following conditions hold:
	\begin{itemize}
		\item[(a)] $Q$ is identifiable in $G$; or
		\item[(b)] There exists $Z' \subseteq Z$ such that the following conditions hold,
		\begin{itemize}
			\item[(i)] $X$ intercepts all directed paths from $Z'$ to $Y$, and 
			\item[(ii)] $Q$ is identifiable in $G_{\overline{Z'}}$.
		\end{itemize}
	\end{itemize}
\end{theorem}

Condition \emph{(a)} is the base case for when standard identifiability is reached. Whenever this is not the case, condition \emph{(b:i)} requires that all directed paths from $Z$ to $Y$ are blocked by $X$. This means that $Z$ has no direct effect on $Y$,  which by the do-calculus implies $P(y | do(x)) = P(y | do(x, z))$; i.e., the effect of $X$ on $Y$ is the same as the effect of $X, Z$ on $Y$. Condition \emph{(b:ii)} notes that manipulation of $Z$ leads to the post-intervention graph $G_{\overline{Z}}$, in which all incoming arrows into $Z$ are deleted. If the effect of $X$ can then be identified in this graph, by the removal of $do(x)$ in the expression, then $\mathpzc{z}$-identification is ascertained. 

For example, recall that in Figure \ref{fig_zid2} the effect of $X$ on $Y$ is not identifiable from $P(v)$. If experimental data over $Z$ is available, i.e., $P(v | do(z))$, then Theorem \ref{th_zid2} can be applied. Note that all the directed paths from $Z$ to $Y$ are blocked by $X$, which satisfies condition \emph{(b:i)}. It is also the case that  in the graph $G_{\overline{Z}}$  (see Figure \ref{fig_zid3}), the set $\{W_1, W_2\}$ is backdoor admissible (by Theorem \ref{adjustment_formula}), which in turn satisfies condition \emph{(b:ii)}. After all, the effect $P(Y=y|do(X=x))$ is identifiable and  given by the expression 

\begin{equation}
\sum_{w_1, w_2} P(Y=y|do(Z=z), X = x, w_1, w_2)P(w_1, w_2 | do(Z=z)).
\end{equation}
As in the observational case, researchers are not required to engage in these derivations by hand, since fully  automated algorithms exist for $\mathpzc{z}$-identification and its generalizations (see \citealp{Bareinboim2012}; and \citealp{Lee:etal19}, for a survey of the latest results).  

$\mathpzc{Z}$-identification exploits experimental variation in a surrogate variable that causally affects the treatment of interest. It thus bears close resemblance to instrumental variable (IV) estimation (\citealp{Wright1928}). The two are not equivalent though. Take the canonical IV setting with an exogenous instrument depicted in Figure \ref{fig_zid4}. In contrast to Figure \ref{fig_zid1}, there is an unobserved confounder directly connecting treatment and outcome. As a result, $P(y|do(x))$ is not $\mathpzc{zID}$ in this graph, because the bidirected arc between $X$ and $Y$ violates condition (\emph{b:ii}) of Theorem \ref{th_zid2}.\footnote{Theorem \ref{th_zid2} is only sufficient, but not necessary. Nonetheless, $\mathpzc{z}$-identification can be proved to be impossible for the graph in Figure \ref{fig_zid4}, following the  general treatment developed in \citet{Lee:etal19}.}

The fact that $P(y|do(x))$ remains unidentifiable in Figure \ref{fig_zid4} is not very surprising, however. It is a well-known result that point identification of the canonical IV estimator is not possible in the nonparametric case (\citealp{Manski1990,Balke1995}). Introducing additional functional form restrictions, such as monotonicity or linearity,  would allow one to identify a \emph{local average treatment effect} for the latent subgroup of compliers (\citealp{Imbens1994}). $\mathpzc{Z}$-identification, by contrast, leverages the fully nonparametric nature of the order relations expressed in causal diagrams. If a query is $\mathpzc{zID}$, the entire post-interventional distribution, including the average treatment effect, is computable from data. Moreover, the solution concepts provided by Theorems \ref{th_zid1} and \ref{th_zid2} are applicable for arbitrary graphs, beyond specific settings such as Figure \ref{fig_zid4}.\footnote{
There exist more refined strategies to identify effects beyond this graph and IVs, including  \citet{Brito2002,chen2016identification,chen2017identification,kumor2019efficient,kumor2020auxiliary}.} Therefore, we consider $\mathpzc{z}$-identification to be an attractive generalization of the IV strategy in fully nonparametric settings. 

   %%%%%%%%%%%%%%%%%%%%%%%%%%%%%%%%%%%%%%%%%%%%%%
\section{Sample Selection Bias}
\label{sec_selection}
%%%%%%%%%%%%%%%%%%%%%%%%%%%%%%%%%%%%%%%%%%%%%%

\setcounter{equation}{0}

The previous section discussed strategies to control for confounding bias, which is the result of nonrandom assignment into treatment. Beyond that, researchers often encounter another source of bias in applied empirical work that stems from preferential selection of units into the data pool. Sample selection poses a serious threat to both statistical as well as causal inference, because it jeopardizes the representativeness of the data for the underlying population. A seminal discussion of this problem in an economic context is given by \citet{Heckman1976,Heckman1979}. He estimates a model of female labor supply in a sample of 2,253 working women interviewed in 1967. The challenge to valid inference in this setting arises due to the fact that market wages are only observable for women who choose to work. His model is described as follows
\begin{align}
s_i &\leftarrow \quad \mathbbm{1}[Z^{'}_{i} \delta - \eta_i > 0], \label{eq_sample_selection1} \\[0.3cm]
y_i &\leftarrow
\begin{cases}
x_i \beta + Z^{'}_{i} \gamma + \varepsilon_i \qquad \text{if } s_i=1,\\ 
\text{unobserved} \hspace{1.25cm} \text{if } s_i=0.
\end{cases} 
\label{eq_sample_selection2} 
\end{align}
Equation (\ref{eq_sample_selection1}) characterizes the sampling mechanism. Wages $y_i$ for an individual $i$ are only observed if $(Z^{'}_{i} \delta - \eta_i)$ attains a value larger than zero, which is captured by the selection indicator $s_i$. Economically, this expresses the idea that individuals will choose to remain unemployed if the market wage they are able to attain (determined by the vector of socioeconomic characteristics $Z_i$) does not exceed their reservation level $\eta_i$. Systematic bias in the coefficient of interest $\beta$ for hours worked $x_i$ can then arise if reservation wages are correlated with unobservables in the market wage equation (\ref{eq_sample_selection2}); that is, if $Corr(\eta_i, \varepsilon_i) \neq 0$.

Similar cases of sample selection are widespread in economics. Examples are discussed by \citet{Levitt2000}, who estimate the effectiveness of seatbelts and airbags in a sample of fatal crashes, and by \citet{Ihlanfeldt1986}, who note the difficulty of assessing the determinants of house prices when using data on recently sold homes. \citet{Knox2019} point out another illustrative case.\footnote{See \citet{Durlauf2020} for a similar argument.} They critique studies which attempt to estimate the extent of racial bias in policing using administrative data (\citealp{Fryer2018}). Problematic in this context is that individuals only appear in such records if police officers decide to stop and interrogate them in the first place. If this stopping decision is itself causally affected by minority status, sample selection bias might arise, since the data is not a representative sample of the overall population anymore.

\begin{figure}[thp]
	\centering
	\subcaptionbox{\label{fig_heckman}}[.36\linewidth]{
		\begin{tikzpicture}[>=triangle 45, font=\footnotesize]
    		\node[fill,circle,inner sep=0pt,minimum size=5pt,label={below:{X}}] (X) at (0,0) {};
    		\node[fill,circle,inner sep=0pt,minimum size=5pt,label={below:{Y}}] (Y) at (3,0) {};
    		\node[fill,circle,inner sep=0pt,minimum size=5pt,label={above:{Z}}] (Z) at (1,1.2) {};
    		\node[draw,circle,inner sep=0pt,double,double distance=0.3mm,minimum size=5pt,label={above:{S}}] (S) at (2.5,1.8) {};
    		\draw[->,shorten >= 1pt] (X)--(Y);
    		\draw[->,shorten >= 1pt] (Z)--(X);
    		\draw[->,shorten >= 1pt] (Z)--(Y);
    		\draw[->,shorten >= 2pt] (Z)--(S);
    		\draw[<->,dashed,shorten >= 2pt] (Y) to[bend right=30] (S);
		\end{tikzpicture}
	}
	\subcaptionbox{\label{fig_simple_slection}}[.26\linewidth]{
		\begin{tikzpicture}[>=triangle 45, font=\footnotesize]
    		\node[fill,circle,inner sep=0pt,minimum size=5pt,label={below:{X}}] (X) at (0,0) {};
    		\node[fill,circle,inner sep=0pt,minimum size=5pt,label={below:{Y}}] (Y) at (2.5,0) {};
    		\node[draw,circle,inner sep=0pt,double,double distance=0.3mm,minimum size=5pt,label={above:{S}}] (S) at (1.25,1) {};
    		\draw[->,shorten >= 1pt] (X)--(Y);
    		\draw[->,shorten >= 2pt] (X)--(S);
		\end{tikzpicture}
	}
	\subcaptionbox{\label{fig_heckman2}}[.36\linewidth]{
		\begin{tikzpicture}[>=triangle 45, font=\footnotesize]
    		\node[fill,circle,inner sep=0pt,minimum size=5pt,label={below:{X}}] (X) at (0,0) {};
    		\node[fill,circle,inner sep=0pt,minimum size=5pt,label={below:{Y}}] (Y) at (3,0) {};
    		\node[fill,circle,inner sep=0pt,minimum size=5pt,label={above:{Z}}] (Z) at (1,1.2) {};
    		\node[fill,circle,inner sep=0pt,minimum size=5pt,label={right:{W}}] (W) at (3.1,1) {};
    		\node[draw,circle,inner sep=0pt,double,double distance=0.3mm,minimum size=5pt,label={above:{S}}] (S) at (2.5,1.8) {};
    		\draw[->,shorten >= 1pt] (X)--(Y);
    		\draw[->,shorten >= 1pt] (Z)--(X);
    		\draw[->,shorten >= 1pt] (Z)--(Y);
    		\draw[->,shorten >= 2pt] (W)--(S);
    		\draw[->,shorten >= 2pt] (W)--(Y);
    		\draw[->,shorten >= 2pt] (Z)--(S);
		\end{tikzpicture}
	}
	\caption{Examples of selection diagrams.}
	\label{fig5}
\end{figure} 

In causal diagrams, cases of sample selection can be captured by explicitly accounting for the sampling selection mechanism. We will realize this goal by adding a new special variable called $S$ to the graph. This variable will take on two values: one, if a unit is part of the sample, and zero otherwise. If endogenous variables in the analysis affect the sampling probabilities, we will add an arrow from these variables to $S$, which will constitute the specification of the selection mechanism (\citealp{Bareinboim2012a}).\footnote{We will consider the case here where the sample selection nodes are only allowed to have incoming arrows, but will not emit arrows themselves.} Figure \ref{fig_heckman} depicts a DAG for the female labor supply example that has been augmented by such a selection node; the resulting graph is denoted by $G_S$. An individual's socioeconomic characteristics $Z$ determine inclusion in the sampling pool and the bidirected dashed arc between $S$ and $Y$ indicates the presence of unobserved confounders that are the source of the error correlation in the model.

Simultaneously controlling for confounding and selection biases introduces a new challenge to the do-calculus. Not only is it necessary to transform interventional distributions into do-free expressions, but the probabilities that make up these expressions now also need to be conditional on $S=1$, because that is all the analyst is able to observe. This additional restriction explains why dealing with selection bias is such a hard problem in practice. At the same time, the literature on recovering causal effects from selection-biased data (\citealp{Bareinboim2012a,Bareinboim2014,Bareinboim2015}) aims to preserve the fully nonparametric nature of causal graphs also in this task. It refrains from introducing functional form assumptions (such as monotonicity or joint normality) related to the selection-propensity score $P(s_i|pa_i)$, as well as \emph{a priori} assuming ignorability of the selection mechanism, which are the approaches most commonly taken in econometrics (\citealp{Angrist1997}.) Nevertheless, even with such a limited set of assumptions as a starting point, several positive results for the recoverability of causal effects from selection bias can be derived.

\subsection{Recoverability of conditional distributions}
\label{sec_recover_cond_dist}

As a first step to make progress, \citet{Bareinboim2014} provide a complete condition for recovering conditional probabilities that do not yet contain a do-operator.

\begin{theorem} 
	\label{th_recoverability}
	(\citealp{Bareinboim2014}) The conditional distribution $P(y|t)$ is recoverable from $G_S$ (as $P(y|t, S=1)$) if and only if $(Y \independent S | T)$.
\end{theorem}
Sufficiency of this condition follows immediately. However, its necessity is less obvious and implies that if $Y$ is not d-separated from $S$ in $G_S$, its conditional distribution will not be recoverable. Combining Theorem \ref{th_recoverability} with do-calculus suggests a straightforward strategy for also recovering do-expressions from selection bias (\citealp{Bareinboim2015}).

\begin{corollary} 
	\label{co_recoverability}
	(\citealp{Bareinboim2015}) The causal effect $Q=P(y|do(x))$ is recoverable from selection-biased data (i.e., $P(v|S=1)$) if using the rules of the do-calculus, Q is reducible to an expression in which no do-operator appears, and recoverability is determined by Theorem \ref{th_recoverability}.
\end{corollary}

Take Figure \ref{fig_simple_slection} as an example. Here, the relationship between $X$ and $Y$ is unconfounded and, therefore, $P(y|do(x)) = P(y|x)$ holds. Moreover, since $S$ and $Y$ are d-separated by $X$, we find the causal effect to be recoverable and given by  $P(y|x,S=1)$. An immediate consequence of Theorem \ref{th_recoverability} is that a causal effect will not be recoverable if $Y$ is directly connected to $S$ via an edge in the graph. Thus, without invoking stronger functional form assumptions there is no possibility to control for selection bias in the female labour supply model of Figure \ref{fig_heckman}.

Selection-biased data complicate identification in observational studies because confounding and selection need to be addressed simultaneously. An example is given by the graph in Figure \ref{fig_selection_example}. Consider the case of a group of entrepreneurs who are looking to crowdfund their business ideas. A researcher is interested in the effect of campaign success $X$ on venture growth $Y$. Idea quality is captured by $Q$, which affects the quality of the crowdfunding campaign, $C$ (possibly multivalued), that is presented to potential investors on a digital platform. To increase their chances of getting funded, some entrepreneurs take part in a training workshop. This decision, $W$, is not necessarily random, which is reflected by the bidirected dashed arc between $X$ and $W$ in the graph. In this example, a problem of selection arises, because the sample that is available to the researcher contains disporportionally many workshop participants, for whom contact addresses were most easily obtainable ($W \rightarrow S$). Finally, $Z$ denotes a set of other confounders that the researcher wishes to control for.\footnote{Figure \ref{fig_selection_example} encodes the assumption that the effect of campaign quality on venture growth is negligible, e.g., because the research question concerns equity crowdfunding in a business-to-business market. Entrepreneurs might additionally use crowdfunding campaigns as an advertising opportunity for their products. Insofar as this gives rise to a direct effect $C \rightarrow Y$, recoverability should be determined with this added assumption.}

Without selection bias, the researcher would have the choice between two backdoor admissible adjustment sets: $\{Z,W,C\}$ and $\{Z,Q\}$. However, with preferential selection into the sample, recoverability can only be obtained with the latter. That is, because in the adjustment formula (Equation \ref{adjustment_formula}),  the prior distribution of the adjustment set needs to be recovered as well, and $\{Z,Q\}$ is the only conditioning set that is  marginally d-separated from $S$ (readers are encouraged to check). Thus, following the strategy dictated by Corollary \ref{co_recoverability}, the estimable backdoor adjustment expression in this example is
\begin{equation}
P(y|do(x)) = \sum_{z,q} P(y|x,z,q,S=1)P(z,q|S=1).
\end{equation}

\begin{figure}[thp]
	\centering
	\subcaptionbox{\label{fig_selection_example}}[.45\linewidth]{
		\begin{tikzpicture}[>=triangle 45, font=\footnotesize]
		\node[fill,circle,inner sep=0pt,minimum size=5pt,label={below:{X}}] (X) at (0,0) {};
		\node[fill,circle,inner sep=0pt,minimum size=5pt,label={below:{Y}}] (Y) at (3.6,0) {};
		\node[fill,circle,inner sep=0pt,minimum size=5pt,label={above:{Q}}] (Q) at (3.6,2) {};
		\node[fill,circle,inner sep=0pt,minimum size=5pt,label={below:{C}}] (C) at (1.8,1) {};
		\node[fill,circle,inner sep=0pt,minimum size=5pt,label={above:{W}}] (W) at (0,2) {};
		\node[fill,circle,inner sep=0pt,minimum size=5pt,label={above:{Z}}] (Z) at (2,-0.8) {};
		\node[draw,circle,inner sep=0pt,double,double distance=0.3mm,minimum size=5pt,label={above:{S}}] (S) at (1.4,2) {};
		\draw[->,shorten >= 1pt] (X)--(Y);
		\draw[->,shorten >= 1pt] (Q)--(Y);
		\draw[->,shorten >= 1pt] (Q)--(C);
		\draw[->,shorten >= 2pt] (C)--(X);
		\draw[->,shorten >= 2pt] (W)--(X);
		\draw[->,shorten >= 2pt] (W)--(C);
		\draw[->,shorten >= 2pt] (W)--(S);
		\draw[->,shorten >= 2pt] (Z)--(X);
		\draw[->,shorten >= 2pt] (Z)--(Y);
		\draw[<->,dashed,shorten >= 1pt] (W) to[bend right=60] (X);
		\draw[<->,dashed,shorten >= 1pt] (Y) to[bend right=60] (Q);
		\end{tikzpicture}
	}
	\subcaptionbox{\label{fig_selection_algorithm}}[.45\linewidth]{
		\begin{tikzpicture}[>=triangle 45, font=\footnotesize]
		\node[fill,circle,inner sep=0pt,minimum size=5pt,label={left:{Z}}] (Z) at (0,0) {};
		\node[fill,circle,inner sep=0pt,minimum size=5pt,label={above:{W}}] (W) at (1,1.2) {};
		\node[fill,circle,inner sep=0pt,minimum size=5pt,label={below:{X}}] (X) at (2,0) {};
		\node[fill,circle,inner sep=0pt,minimum size=5pt,label={right:{Y}}] (Y) at (4,0) {};
		\node[draw,circle,inner sep=0pt,double,double distance=0.3mm,minimum size=5pt,label={above:{S}}] (S) at (2.8,0.8) {};
		\draw[->,shorten >= 1pt] (X)--(Y);
		\draw[->,shorten >= 1pt] (Z)--(X);
		\draw[->,shorten >= 1pt] (W)--(X);
		\draw[->,shorten >= 1pt] (W)--(Z);
		\draw[->,shorten >= 2pt] (W)--(S);
		\draw[<->,dashed,shorten >= 1pt] (Z) to[bend right=45] (Y);
		\end{tikzpicture}
	}
	\caption{More challenging examples for recovering from selection bias.}
	\label{fig6}
\end{figure}

\subsection{A general solution for recovering from selection bias}
\label{sec_selection_general}

It is important to note that although Theorem \ref{th_recoverability} provides a necessary condition for recovering conditional probabilities, the same does not hold for Corollary \ref{co_recoverability} with respect to do-expressions. This is exemplified by the graph in Figure \ref{fig_selection_algorithm}. Due to unobserved confounders between $Z$ and $Y$, and the fact that $Z$ is a collider on the path $X \leftarrow W \rightarrow Z \dashleftarrow\dashrightarrow Y$, identification via the backdoor criterion would require to adjust for both $Z$ and $W$, which will close all the backdoor paths. However, $\{Z,W\}$ is not d-separable from $S$ ($W$ has a direct arrow to $S$), and an attempt to apply Corollary \ref{co_recoverability} will thus fail. Nevertheless, and perhaps surprisingly, $P(y|do(x))$ can still be recovered in Figure \ref{fig_selection_algorithm} with the help of do-calculus using a slightly more sophisticated approach.\footnote{The following do-calculus derivations are shown in more detail, with corresponding subgraphs depicted alongside, in Appendix \ref{app_do-calculus2}.} To witness, note that $(S,W \independent Y)$ in $G_{\overline{X}}$, i.e., the resulting graph when all incoming arrows in $X$ are deleted (see Section \ref{sec_do-calculus}). Then, according to the first rule of do-calculus 
\begin{align}
P(y|do(x)) &= P(y|do(x), w, S=1),  \\[0.3cm]
&= \sum_z P(y|do(x), z, w, S=1) P(z|do(x), w, S=1), \label{eq_s-do-calculus} 
\end{align}
where the second line follows by conditioning on $Z$. Applying rule 2 of do-calculus, since $(Y \independent X | W,Z,S)$ in $G_{\underline{X}}$, the do-operator can be removed in the first term of equation \ref{eq_s-do-calculus}
\begin{equation}
= \sum_z P(y|x, z, w, S=1) P(z|do(x), w, S=1).
\end{equation}
Finally, since $(Z \independent X | W,S)$ in $G_{\overline{X(W)}}$, rule 3 of the calculus allows us to remove the $do(x)$ from the second term, such that
\begin{equation}
P(y|do(x)) = \sum_z P(y|x, z, w, S=1) P(z| w, S=1).
\end{equation}
Note that the quantities in the final expression of $P(y|do(x))$ do not involve any do-operator, since the data are observational and always contain $S=1$, given that the samples were selected preferentially. Taken together, this ensures recoverability of the target interventional distribution.

\citet{Bareinboim2015} provide algorithmic criteria for recovering interventional distributions (i.e., containing $do(x)$-operators) in arbitrary causal graphs. They permit full automation of derivations such as the one just performed. Recently, this algorithm was also proved complete for the recovery task by \citet{Correa2019}. 

\subsection{Combining biased and unbiased data}
\label{sec_unbiased_data}

Another promising strategy for recovering causal quantities from sample selection is when biased and unbiased data sources are combined. For example, the distributions of important socioeconomic variables can often be measured without bias, e.g., from population-level statistics. Interestingly, this is the case in the original Heckman selection model, where number of children, household assets, husband's wage, labour market experience, and education are observed both for working and non-working women (\citealp{Heckman1976}).  To illustrate how this can facilitate recoverability, we revisit the example from Figure \ref{fig_heckman}, but now assume that the common parent node of wages $Y$ and the selection node $S$ is observable as $W$ (see Figure \ref{fig_heckman2}, which is the same as Figure \ref{fig_heckman} but for the replacement of the bidirected arrow with the observed $W$). If that is the case, conditioning on the set $\{Z,W\}$ closes all backdoor paths between $X$ and $Y$ and simultaneously d-separates $Y$ from $S$. From the backdoor adjustment formula discussed above (Theorem \ref{th_backdoor_adjustment_formula}), we can thus derive
\begin{align}
P(y|do(x)) &= \sum_{z,w} P(y|x, z, w) P(z,w), \\[0.3cm]
&= \sum_{z,w} P(y|x,z,w,S=1) P(z,w), \label{eq_s-backdoor} 
\end{align}
where the second line follows from Theorem \ref{th_recoverability}, since $(Y \independent S | Z, W)$. As $P(z,w)$ cannot be recovered from selection bias, Corollary \ref{co_recoverability} is not applicable.  However, if in addition to the selection-biased data, unbiased measurements of $P(z,w)$ are available (e.g., from census data), equation (\ref{eq_s-backdoor}) becomes estimable. 

\citet{Bareinboim2014} leverage this idea and present the following generalization of the backdoor criterion, which can be invoked if a subset $Z$ of the data is measured without bias.
\begin{definition}
	\label{def_s-backdoor}
	(Selection backdoor criterion; \citealp{Bareinboim2014}) Let a set $Z$ of variables be partitioned into $Z^{+} \cup Z^{-}$ such that $Z^{+}$ contains all non-descendants of $X$ and $Z^{-}$ the descendants of $X$, and let $G_S$ stand for the graph that includes sampling mechanism $S$. $Z$ is said to satisfy the selection backdoor criterion (s-backdoor, for short) if it satisfies the following conditions: 
	\setcounter{bean}{0}
	\begin{list}
		{\arabic{bean}.}{\usecounter{bean}}
		\item $Z^{+}$ blocks all backdoor paths from $X$ to $Y$ in $G_S$;
		\item $X$ and $Z^{+}$ block all paths between $Z^{-}$ and $Y$ in $G_S$, namely, $(Z^{-} \independent Y | X, Z^{+})$;
		\item $X$ and $Z$ block all paths between $S$ and $Y$ in $G_S$, namely, $(Y \independent S | X, Z)$; and
		\item $Z$ and $Z \cup \{X,Y\}$ are measured in the unbiased and biased studies, respectively.
	\end{list}
\end{definition}
The following theorem can then be proved. 
\begin{theorem} 
	\label{th_s-backdoor}
	(\citealp{Bareinboim2014}) If $Z$ is s-backdoor admissible, then causal effects are identified by
	\begin{equation}
	P(y|do(x)) = \sum_{z} P(y|x,z,S=1) P(z).
	\end{equation} 
\end{theorem}

The s-backdoor criterion is a sufficient condition for generalized adjustment, which is able to deal with confounding and selection bias simultaneously. \citet{Correa2018} substantially extend this line of work by presenting conditions that are both necessary \emph{and} sufficient. Furthermore, \citet{Correa2019} provide a sound algorithm for recovering causal effects from a mix of biased and unbiased data in causal graphs that are arbitrary in size and shape.

   %%%%%%%%%%%%%%%%%%%%%%%%%%%%%%%%%%%%%%%%%%%%%%
\section{Transportability of Causal Knowledge}
\label{sec_transportability}
%%%%%%%%%%%%%%%%%%%%%%%%%%%%%%%%%%%%%%%%%%%%%%

\setcounter{equation}{0}

Extrapolating causal knowledge across domains is a fundamental problem in causal inference. Experiments are usually conducted in different contexts than those in which the lessons drawn from them are supposed to be applied. Expecting experimental results to hold across populations may be fallacious, however, if domains differ structurally in important ways. \citet{Duflo2008} allude to this problem in a development economics context when asking: \emph{``If a program worked for poor rural women in Africa, will it work for middle-income urban men in South Asia?"}. In this section, we discuss the conditions under which a transfer of causal knowledge across structurally heterogeneous domains is valid. This issue is known under the rubric of \emph{``transportability"} in the artificial intelligence literature, while social scientists usually refer to it as \emph{``external validity"} (\citealp{Shadish2002,Bareinboim2013,Pearl2014}). \citet{Nakamura2018} discuss the challenge of external validity from a macroeconomic perspective and come to the conclusion that \emph{``even very cleanly identified monetary and fiscal natural experiments give us, at best, only a partial assessment of how future monetary and fiscal policy actions---which may differ in important ways from those in the past---will affect the economy."} Causal diagrams, in conjunction with do-calculus, allow to formally address these kinds of concerns in a principled, general, and efficient way, eliciting the assumptions needed to analyze these settings and making precise how much can actually be learned from experiments across different domains.

In practice, it is often implicitly assumed that an experimental result obtained in a population $\Pi$ provides at least a good approximation for the impact of the same intervention in other settings. This assumption is made for convenience, because it allows to use results from $\Pi$ for policy decisions in a different population $\Pi^*$. However, such kind of \emph{direct transportability}, which we formally define in the following, is likely to be violated in many empirical settings.

\begin{definition}
	\label{def_direct_transportability} 
	(Direct Transportability; \citealp{Pearl2011}) A causal relation $R$ is said to be directly transportable from $\Pi$ to $\Pi^*$, if $R(\Pi^*) = R(\Pi)$.
\end{definition}

For an example, consider the study by \citet{Banerjee2007} that analyzes the effects of a remedial education program in two major cities in Western India: Mumbai and Vadodara. The randomized intervention provided schools with an extra teacher for tutoring children in the third and fourth grades, who had been lagging behind their peers. The program showed substantial positive effects on children's academic achievements, at least in the short-run. Interestingly, however, while treatment effects on mathematics scores were similar in both cities, the effect on language proficiency was weaker in Mumbai compared to Vadodara. The authors explain this finding by higher baseline reading skills in Mumbai, where families were on average wealthier and schools were better equipped. By contrast, baseline skill levels in mathematics did not differ significantly. The remedial education program, which targeted only the most basic competencies in the curriculum, was therefore equally effective for mathematics skills.

\begin{figure}[thp]
	\centering
	\subcaptionbox{\label{fig_transportability1}}[.32\linewidth]{
		\begin{tikzpicture}[>=triangle 45, font=\footnotesize]
		\node[fill,rectangle,inner sep=0pt,minimum size=5pt,label={right:{S}}] (S) at (2.7,2.7) {};
		\node[fill,circle,inner sep=0pt,minimum size=5pt,label={below:{X}}] (X) at (0,0) {};
		\node[fill,circle,inner sep=0pt,minimum size=5pt,label={below:{Y}}] (Y) at (4,0) {};
		\node[fill,circle,inner sep=0pt,minimum size=5pt,label={above:{Z}}] (Z) at (2,2) {};
		\draw[->,shorten >= 1pt] (X)--(Y);
		\draw[->,shorten >= 1pt] (Z)--(X);
		\draw[->,shorten >= 1pt] (Z)--(Y);
		\draw[->,shorten >= 1pt] (S)--(Z);
		\draw[<->,dashed,shorten >= 1pt] (X) to[bend left=45] (Z);
		\draw[<->,dashed,shorten >= 1pt] (X) to[bend left=30] (Y);
		\end{tikzpicture}
	}
	\subcaptionbox{\label{fig_transportability2}}[.32\linewidth]{
		\begin{tikzpicture}[>=triangle 45, font=\footnotesize]
		\node[fill,rectangle,inner sep=0pt,minimum size=5pt,label={right:{S}}] (S) at (0,1) {};
		\node[fill,circle,inner sep=0pt,minimum size=5pt,label={[xshift=-2]below:{X}}] (X) at (0,0) {};
		\node[fill,circle,inner sep=0pt,minimum size=5pt,label={[xshift=2]below:{Y}}] (Y) at (3,0) {};
		\draw[->,shorten >= 1pt] (X)--(Y);
		\draw[->,shorten >= 1pt] (S)--(X);
		\draw[<->,dashed,shorten >= 1pt] (X) to[bend left=50] (Y);
		\end{tikzpicture}
	}
	\subcaptionbox{\label{fig_transportability3}}[.32\linewidth]{
		\begin{tikzpicture}[>=triangle 45, font=\footnotesize]
		\node[fill,rectangle,inner sep=0pt,minimum size=5pt,label={right:{S}}] (S) at (2,3) {};
		\node[fill,circle,inner sep=0pt,minimum size=5pt,label={below:{X}}] (X) at (0,0) {};
		\node[fill,circle,inner sep=0pt,minimum size=5pt,label={below:{Y}}] (Y) at (4,0) {};
		\node[fill,circle,inner sep=0pt,minimum size=5pt,label={[xshift=-7,yshift=-2]above:{Z}}] (Z) at (2,2) {};
		\draw[->,shorten >= 1pt] (X)--(Y);
		\draw[->,shorten >= 1pt] (Z)--(X);
		\draw[->,shorten >= 1pt] (Z)--(Y);
		\draw[->,shorten >= 1pt] (S)--(Z);
		\draw[<->,dashed,shorten >= 1pt] (X) to[bend left=45] (Z);
		\draw[<->,dashed,shorten >= 1pt] (Z) to[bend left=45] (Y);
		\draw[<->,dashed,shorten >= 1pt] (X) to[bend left=30] (Y);
		\end{tikzpicture}
	}
	\caption{Examples of selection diagrams for the transportability task.}
	\label{fig7}
\end{figure}

The graph in Figure \ref{fig_transportability1} provides a graphical representation of the setting in \citet{Banerjee2007}. Assume that we want to generalize experimental results from a trial conducted in Vadodara ($\Pi$) to the population in Mumbai ($\Pi^*$). However, we are aware of the fact that income levels of families $Z$, which are an important determinant of children's academic achievements $Y$, are higher in Mumbai. In a causal diagram, we can incorporate this domain knowledge ex-ante by adding a set of \emph{selection nodes} $S$ that indicate where both populations under study differ, either in the distribution of background factors $P(u)$ or due to divergent causal mechanisms $f_i$. These $S$-nodes thus locate the sources of structural discrepancies across domains that threaten transportability. Switching between two populations $\Pi$ and $\Pi^*$ is then captured by conditioning on different values of $S$.\footnote{For clarity, $S$-nodes invoked for transportability are depicted by squares ($\blacksquare$), in order to distinguish them from the selection bias case. Also note that now $S$ is emitting arrows, whereas selection nodes indicating preferential inclusion into the sample only receive arrows.} Next, we define the joint graphical representation of the corresponding structural models in the source and target populations, which is required to establish transportability. 

\begin{definition}
	\label{def_selection_diagram}
	(Selection Diagram; \citealp{Pearl2011}) Let $\langle M, M^* \rangle$ be a pair of structural causal models (see Definition \ref{def_scm}) relative to domains $\langle \Pi, \Pi^* \rangle$, sharing a causal diagram G. $\langle M, M^* \rangle$ is said to induce a selection diagram $D$ if $D$ is constructed as follows: (a) every edge in $G$ is also an edge in $D$; and (b) $D$ contains an extra edge $S_i \rightarrow V_i$ whenever there might exist a discrepancy $f_i \neq f_i^*$ or $P(U_i) \neq P^*(U_i)$ between $M$ and $M^*$.
\end{definition}

The absence of an $S$-node in the selection diagram represents the assumption that the causal mechanism, which assigns values to the respective variable, is the same in both populations. In the extreme case, one could add $S$-nodes to all nodes in the graph, to express the notion that the two populations are maximally structurally heterogeneous (i.e., there are no structural invariances). Obviously, this would undermine any hope for information exchange across domains though.

Equipped with the definition of a selection diagram, we can state the following theorem, which allows to transport experimental results obtained in a source $\Pi$ to another target domain $\Pi^*$, where only passive observation is possible.\footnote{Following Definition \ref{def_selection_diagram}, both domains $\Pi$ and $\Pi^*$ share the same causal diagram $G$. Consequently, if a causal query Q is identifiable with observational data alone in the source domain $\Pi$ (i.e., no experimental knowledge is necessary), it will also be identifiable in the target domain $\Pi^*$, and $Q$ will thus be \emph{trivially transportable} (\citealp{Pearl2011}). \citet{Pearl2011} discuss \emph{observational transportability} of a statistical query of the form $P(y|x)$ (e.g., a classifier) from a source domain to a target domain, where only a subset of the variables in the selection diagram are observed. Thus, statistical transportability permits the analyst to save on data collection costs. Later on, \citet{Correa2019b} developed a complete algorithm for this task. We will not further pursue this topic in what follows and refer the interested reader to the respective papers.}

\begin{theorem}
	\label{th_transportability1}
	(\citealp{Pearl2011}) Let $D$ be the selection diagram characterizing two populations, $\Pi$ and $\Pi^*$, and $S$ the set of selection variables in $D$. The strata-specific causal effect $P^*(y|do(x),z)$ is transportable from $\Pi$ to $\Pi^*$ if $Z$ d-separates $Y$ from $S$ in the $X$-manipulated version of $D$, that is, $Z$ satisfies $(Y \independent S|Z,X)_{D_{\overline{X}}}$.
\end{theorem}

Note that $D_{\overline{X}}$ refers to the post-intervention graph, in which all incoming arrows into $X$ are deleted (see Section \ref{sec_do-calculus}). D-separation between $S$-nodes and the outcome variable $Y$  can be achieved by adjusting for a conditioning set $T$, as the following definition formalizes.

\begin{definition}
	\label{def_s-admissibility}
	(S-admissibility; \citealp{Pearl2011}) A set $T$ of variables satisfying $(Y \independent S |T)$ in $D_{\overline{X}}$ will be called s-admissible (with respect to the causal effect of $X$ on $Y$).
\end{definition}

Syntactically, this result is somewhat similar to the selection bias case (see Theorem \ref{th_recoverability}), where the selection indicator was likewise required to be d-separated from $Y$ by a set $T$ (\citealp{Pearl2015a}). Semantically, this separation of an $S$-node indicates that the target distribution is insensitive to the structural disparities represented in the selection diagram, and, therefore, the effects are invariant across populations. Looking at the selection diagram in Figure \ref{fig_transportability1}, we note that the set $\{Z\}$ d-separates $S$ and $Y$ in $D_{\overline{X}}$ (i.e., when $X$ is experimentally manipulated). It therefore satisfies s-admissibility. 

By applying the rules of do-calculus, we can now show that s-admissibility implies transportability across domains.
\begin{align}
P^*(y|do(x)) &= P(y|do(x),s), \\[0.3cm]
&= \sum_z P(y|do(x),z,s) P(z|do(x),s), \\[0.3cm]
&= \sum_z P(y|do(x),z) P(z|s), \\[0.3cm]
&= \sum_z P(y|do(x),z) P^*(z).
\end{align}
The first equation follows from the definition that distributions in the target domain $\Pi^*$ are denoted by conditioning on $S$. The second line follows from conditioning and summing over $Z$. The third line is derived by using the s-admissibility of $Z$ and recognizing the fact that $X$ is a child of $Z$ and, therefore, exerts no causal influence on $Z$ (formally, rule 3 of do-calculus applies). The last line is then just a restatement. 

As long as Figure \ref{fig_transportability1} provides an accurate model for the setting in \citet{Banerjee2007}, the causal effect of the remedial education program in Mumbai can thus be computed by reweighting the stratum-specific causal effect (for every income level of $Z$) obtained in Vadodara by the income distribution $P^*(z)$ in Mumbai. No experimental data for Mumbai is required. This result is stated in its full generality in the following corollary.

\begin{corollary}
	\label{cor_transportability1}
	(\citealp{Pearl2011}) The causal effect $P^*(y|do(x))$ is transportable from $\Pi$ to $\Pi^*$ if there exists a set $Z$ of observed pretreatment covariates that is s-admissible. Moreover, the transport formula is given by the weighting	
	\begin{equation}
		\label{eq_transport}
		P^*(y|do(x)) = \sum_z P(y|do(x),z) P^*(z).
	\end{equation} 
\end{corollary}
It is an immediate consequence of Theorem \ref{th_transportability1} that any $S$ variable that points into $X$ can be ignored. The causal effect $P(y|do(x))$ is thus directly transportable in Figure \ref{fig_transportability2}. The same holds for $S$ nodes that are d-separated by the empty set in $D_{\overline{X}}$.

As a graphical criterion, s-admissibility is easy to check. Without a reference to a causal diagram, however, the intricacies of transportability can be hard to discern. Figure \ref{fig_transportability3} provides a cautionary tale in that regard. Apart from the unobserved confounder between $Z$ and $Y$, it is identical to Figure \ref{fig_transportability1}. Here, however, s-admissibility is violated because conditioning on $Z$ would open up the path $S \rightarrow Z \dashleftarrow\dashrightarrow Y$. It can be shown that transporting $P(y|do(x))$ is impossible in this selection diagram. The example thus illustrates how the absence or presence of one single edge can determine whether transportability is feasible. Recognizing such subtleties by pure introspection, without the reference to an explicit model, would be an extremely difficult undertaking.

The transport formula presented in equation (\ref{eq_transport}) is well known in the econometrics literature (\citealp{Hotz2005,Dehejia2015,Andrews2018}). Most commonly, approaches in this area build on the potential outcomes framework, where s-admissibility is encoded through ignorability relations; i.e., domain heterogeneity $S$ is assumed to be ignorable given pretreatment covariates $X$. While it is hard to judge ignorability statements, we note that this assumption is easily violated in practice, as the example in Figure \ref{fig_transportability3} demonstrates. Causal graphs offer valuable guidance for judging the validity of ignorability assumptions, which is missing in the potential outcomes framework. Furthermore, using the rules of do-calculus, it becomes possible to establish transportability in more general cases that are not covered by Corollary \ref{cor_transportability1}. 

\begin{theorem}
	\label{th_transportability2}
	(\citealp{Pearl2011}) Let $D$ be the selection diagram characterizing two populations, $\Pi$ and $\Pi^*$, and $S$ as set of selection variables in $D$. The relation $R = P^*(y|do(x))$ is transportable from $\Pi$ to $\Pi^*$ if the expression $P(y|do(x),s)$ is reducible, using the rules of do-calculus, to an expression in which $S$ appears only as a conditioning variable in do-free terms.
\end{theorem}

\begin{figure}[thp]
	\centering
	\subcaptionbox{\label{fig_transportability4}}[.35\linewidth]{
		\begin{tikzpicture}[>=triangle 45, font=\footnotesize]
		\node[fill,rectangle,inner sep=0pt,minimum size=5pt,label={right:{S}}] (S) at (2,0.9) {};
		\node[fill,circle,inner sep=0pt,minimum size=5pt,label={[xshift=-2]below:{X}}] (X) at (0,0) {};
		\node[fill,circle,inner sep=0pt,minimum size=5pt,label={[xshift=2]below:{Y}}] (Y) at (4,0) {};
		\node[fill,circle,inner sep=0pt,minimum size=5pt,label={below:{Z}}] (Z) at (2,0) {};
		\draw[->,shorten >= 1pt] (X)--(Z);
		\draw[->,shorten >= 1pt] (Z)--(Y);
		\draw[->,shorten >= 1pt] (S)--(Z);
		\draw[->,shorten >= 1pt] (X) to[bend right=45] (Y);
		\draw[<->,dashed,shorten >= 1pt] (X) to[bend left=75] (Y);
		\end{tikzpicture}
	}
	\subcaptionbox{\label{fig_transportability5}}[.5\linewidth]{
		\begin{tikzpicture}[>=triangle 45, font=\footnotesize]
		\node[fill,rectangle,inner sep=0pt,minimum size=5pt,label={above:{S$^{'}$}}] (S) at (3.85,2.65) {};
		\node[fill,rectangle,inner sep=0pt,minimum size=5pt,label={above:{S}}] (S2) at (1.15,2.65) {};
		\node[fill,circle,inner sep=0pt,minimum size=5pt,label={[xshift=-2]below:{X}}] (X) at (0,0) {};
		\node[fill,circle,inner sep=0pt,minimum size=5pt,label={[xshift=2]below:{Y}}] (Y) at (5,0) {};
		\node[fill,circle,inner sep=0pt,minimum size=5pt,label={[xshift=4]below:{Z}}] (Z) at (2.5,0) {};
		\node[fill,circle,inner sep=0pt,minimum size=5pt,label={[xshift=-6,yshift=-2]above:{W$_1$}}] (W1) at (0.5,2) {};
		\node[fill,circle,inner sep=0pt,minimum size=5pt,label={[xshift=-5,yshift=2]below:{W$_2$}}] (W2) at (1.5,1) {};
		\node[fill,circle,inner sep=0pt,minimum size=5pt,label={[xshift=6,yshift=-2]above:{W$_3$}}] (W3) at (4.5,2) {};
		\draw[->,shorten >= 1pt] (S)--(W3);
		\draw[->,shorten >= 1pt] (S2)--(W1);
		\draw[->,shorten >= 1pt] (X)--(Z);
		\draw[->,shorten >= 1pt] (Z)--(Y);
		\draw[->,shorten >= 1pt] (W1)--(X);
		\draw[->,shorten >= 1pt] (W1)--(W2);
		\draw[->,shorten >= 1pt] (W2)--(Z);
		\draw[->,shorten >= 1pt] (W3)--(Z);
		\draw[->,shorten >= 1pt] (W3)--(Y);
		\draw[<->,dashed,shorten >= 1pt] (X) to[bend right=45] (Y);
		\draw[<->,dashed,shorten >= 1pt] (X) to[bend left=45] (W1);
		\draw[<->,dashed,shorten >= 1pt] (W1) to[bend left=45] (W2);
		\draw[<->,dashed,shorten >= 1pt] (X) to[bend right=40] (Z);
		\end{tikzpicture}
	}
	\caption{Examples of more challenging transportability tasks, including $S$-nodes on post-treatment variables.}
	\label{fig8}
\end{figure}

One such class of models is given when domains differ due to variables that are themselves causally affected by the treatment, as in Figure \ref{fig_transportability4}. Here, the effect of $X$ on $Y$ is partly transmitted by $Z$, and domains differ either according to the distribution of background factors $U_Z$ or the mechanism $f_Z$ that determines $Z$. Such a situation can occur for RCTs in development economics, where the success of a policy is partly dependent on the level of care with which a program is implemented. \citet{Duflo2008} discuss the problem that pilot trials often employ particularly highly qualified program officials. This is difficult to replicate once the program is supposed to be scaled up, which threatens the generalizability of these pilot studies.\footnote{Similarly, \citet{Banerjee2017} discuss how market equilibrium effects can be an obstacle for the generalizability of pilot studies. A large, nationwide experiment may have an effect on wages and prices of nontradable goods such as land, which is likely to be negligible in smaller RCTs. These intermediate variables might be important for the overall outcome of a program and could thus lead to different expected results in a small versus a larger study population.} 

\citet{Gordon2018} provide a similar example from an entirely different context. The effectiveness of advertising campaigns on social media platforms depends on how frequently clients are exposed to the ads. Exposure thus acts as a mediator for the effect of advertising on an outcome of interest, e.g., the click-through rate. And since exposure is determined by user behavior, it cannot easily be controlled by the advertiser. If a social media company running advertising experiments wants to transport results obtained on a desktop version of the platform to users with mobile devices, it will need to take into account that exposure might differ across domains, e.g., due to differences in user demographics.

If post-treatment variables, such as in Figure \ref{fig_transportability4}, are s-admissible, the causal effect of $X$ can be transported as
\begin{align}
	P^*(y|do(x)) &= P(y|do(x),s), \\[0.3cm]
	&= \sum_z P(y|do(x),z,s) P(z|do(x),s), \\[0.3cm]
	&= \sum_z P(y|do(x),z) P^*(z|do(x)), \label{eq_transportability_post-treatment3}
\end{align}
where the last line follows from s-admissibility (\citealp{Pearl2014}). Given equation (\ref{eq_transportability_post-treatment3}), we can see that transportability of $P^*(y|do(x))$ then requires to transform  $P^*(z|do(x))$ into a do-free expression, since by definition no manipulation can be carried out in the target domain. Recognizing that $X$ and $Z$ are unconfounded in Figure \ref{fig_transportability4},  this can be achieved by setting $P^*(z|do(x)) = P^*(z|x)$ (formally, rule 2 of do-calculus applies).

The resulting transport formula, when domains differ according to post-treatment variables, is different from the simple expression in equation (\ref{eq_transport}). It prescribes to reweight the $z$-specific effects by the conditional (instead of the uncoditional) distribution of $Z$ in the target population

\begin{equation}
P^*(y|do(x)) = \sum_z P(y|do(x),z) P^*(z|x).
\end{equation}

Theorem \ref{th_transportability2} was proven to be a necessary and sufficient criterion for transporting causal effect estimates across domains by \citet{Bareinboim2012b}. However, it is only procedural in nature and, therefore, does not specify the sequence of do-calculus steps that need to be taken to arrive at the desired expression. In order to fill this gap, \citet{Bareinboim2013} develop a complete algorithmic solution for carrying out the transformation. The benefits of solving the transportability problem algorithmically become particularly apparent for more complex graphs, such as in Figure \ref{fig_transportability5}, in which the correct transport formula is given by
\begin{align}
P^*(y|do(x)) = \sum_{z,w_2,w_3} P(y|do(x),z,w_2,w_3) P(z|do(x),w_2,w_3) P^*(w_2,w_3).
\end{align}
\noindent Note also that this expression does not contain $W_1$. Applying the transportability algorithm thus helps to decide which measurements are required for transportability and thereby allows to economize on data collection efforts in the target domain. 

\subsection{Transportability with surrogate experiments}

\citet{Bareinboim2013a} combine the idea of transportability with the previously introduced concept of $\mathpzc{z}$-identification, to develop a theory they call \emph{$\mathpzc{z}$-transportability}. Owing to this extension, it becomes possible to not only transfer causal knowledge obtained from direct randomized control trials, but also from the encouragement designs, discussed in Section \ref{sec_z-identifiability}, that rely on surrogate experiments. Researchers are thus given the flexibility to learn from knowledge across domains even in cases when direct manipulation of a treatment would be prohibitively costly, both in the target and in the source domain. 

Remarkably, $\mathpzc{z}$-transportability is a distinct problem and reduces neither to ordinary transportability nor to $\mathpzc{z}$-identifiability. \citet{Bareinboim2013a} demonstrate this fact by presenting examples of causal queries which are $\mathpzc{zID}$ in the source domain $\Pi$, but that may or may not be $\mathpzc{z}$-transportable. Analogous to Theorem \ref{th_transportability2}, the rules of do-calculus can be used to transfer causal knowledge from surrogate experiments in the following way.

\begin{theorem}
	\label{th_z-transportability}
	(\citealp{Bareinboim2013a,Bareinboim2014b}) Let $D$ be the selection diagram characterizing two populations, $\Pi$ and $\Pi^*$, and $S$ be the  set of selection variables in $D$. The relation $R = P^*(y|do(x))$ is $\mathpzc{z}$-transportable from $\Pi$ to $\Pi^*$ in $D$ if and only if the expression $P(y|do(x),s)$ is reducible, using the rules of do-calculus, to an expression in which all do-operators apply to subsets of $Z$, and the $S$-variables are separated from these do-operators.
\end{theorem}

Again, Theorem \ref{th_z-transportability} provides no indication of the sequence of do-calculus steps that need to be taken in order to establish $\mathpzc{z}$-transportability. To this end, \citet{Bareinboim2013a} develop a complete algorithm, which takes the selection diagram $D$ and a list of variables that were manipulated in the source domain as inputs and then returns a transport formula expression whenever such an expression exists.

\subsection{Combining causal knowledge from several heterogeneous source domains}
\label{sec_m-transportability}

Transportability techniques are particularly valuable in situations that allow to combine empirical knowledge from several source domains. \citet{Dehejia2015} consider the case of a policy-maker who is faced with the decision to either learn about a desired treatment effect from extrapolation of an existing experimental evidence base, or to commission a costly new experiment. The challenge in this situation is that previous experiments have possibly been conducted in very different contexts than the one of interest, and underlying populations might be quite heterogeneous. Na\"{i}ve pooling of results is thus likely to fail. Based on the approaches presented in the previous sections, \citet{Bareinboim2013b} introduce the concept of \emph{meta}-transportability (or $\mu$-transportability, for short), which provides a principled solution to this problem.\footnote{Meta-transportability is related to the idea of ``data combination" presented e.g.\ in \citet{Ridder2007}. In this case, however, the goal is to combine causal knowledge from several heterogeneous populations that share at least some causal mechanisms.} 

\begin{figure}[thp]
	\centering
	\subcaptionbox{\label{fig_meta-transportability1}}[.45\linewidth]{
		\begin{tikzpicture}[>=triangle 45, font=\footnotesize]
		\node[fill,rectangle,inner sep=0pt,minimum size=5pt,] (S1) at (0,0.9) {};
		\node[fill,rectangle,inner sep=0pt,minimum size=5pt,] (S2) at (4,0.9) {};
		\node[fill,circle,inner sep=0pt,minimum size=5pt,label={[xshift=-2]below:{X}}] (X) at (0,0) {};
		\node[fill,circle,inner sep=0pt,minimum size=5pt,label={[xshift=2]below:{Y}}] (Y) at (4,0) {};
		\node[fill,circle,inner sep=0pt,minimum size=5pt,label={below:{Z}}] (Z) at (2,0) {};
		\draw[->,shorten >= 1pt] (X)--(Z);
		\draw[->,shorten >= 1pt] (Z)--(Y);
		\draw[->,shorten >= 1pt] (S1)--(X);
		\draw[->,shorten >= 1pt] (S2)--(Y);
		\draw[->,shorten >= 1pt] (X) to[bend right=45] (Y);
		\draw[<->,dashed,shorten >= 1pt] (X) to[bend left=75] (Y);
		\draw[<->,dashed,shorten >= 1pt] (X) to[bend left=45] (Z);
		\end{tikzpicture}
	}
	\subcaptionbox{\label{fig_meta-transportability2}}[.45\linewidth]{
		\begin{tikzpicture}[>=triangle 45, font=\footnotesize]
		\node[fill,rectangle,inner sep=0pt,minimum size=5pt,] (S1) at (0,0.9) {};
		\node[fill,rectangle,inner sep=0pt,minimum size=5pt,] (S2) at (2,0.9) {};
		\node[fill,circle,inner sep=0pt,minimum size=5pt,label={[xshift=-2]below:{X}}] (X) at (0,0) {};
		\node[fill,circle,inner sep=0pt,minimum size=5pt,label={[xshift=2]below:{Y}}] (Y) at (4,0) {};
		\node[fill,circle,inner sep=0pt,minimum size=5pt,label={below:{Z}}] (Z) at (2,0) {};
		\draw[->,shorten >= 1pt] (X)--(Z);
		\draw[->,shorten >= 1pt] (Z)--(Y);
		\draw[->,shorten >= 1pt] (S1)--(X);
		\draw[->,shorten >= 1pt] (S2)--(Z);
		\draw[->,shorten >= 1pt] (X) to[bend right=45] (Y);
		\draw[<->,dashed,shorten >= 1pt] (X) to[bend left=75] (Y);
		\draw[<->,dashed,shorten >= 1pt] (X) to[bend left=45] (Z);
		\end{tikzpicture}
	}
	\caption{Selection diagrams representing two heterogeneous source domains.}
	\label{fig9}
\end{figure}

Let $\mathcal{D} =  \{D_1,\hdots,D_n\}$ be a collection of selection diagrams relative to source domains $\Pi = \{\pi_1,\hdots,\pi_n\}$. An example is given by Figure \ref{fig9}, in which panel (a) depicts the selection diagram that corresponds to source domain $\pi_a$, while panel (b) refers to $\pi_b$. Square nodes indicate where discrepancies between the target domain $\pi^*$ and the source domains arise.\footnote{The causal diagram for the target domain is accordingly obtained by deleting all square nodes from the selection diagrams.} In line with Definition \ref{def_selection_diagram}, these discrepancies can occur due to differences in causal mechanisms as well as background factors related to the the variables that square nodes point into.

Figure \ref{fig9} is a simple extension of a graph that was presented earlier (see Figure \ref{fig_transportability4}). In contrast to before, the unobserved confounder between $X$ and $Z$ (denoted by the dashed bidirected arc $X \dashleftarrow\dashrightarrow Z$), which was added to the diagram, now renders individual transportability impossible.\footnote{The algorithm by \citet{Bareinboim2013} would exit without returning a transport formula expression for both selection diagrams. Intuitively, in panel (a), transportability is prohibited by the selection node pointing directly into $Y$. In (b), $X \dashleftarrow\dashrightarrow Z$ prevents to set $P^*(z|do(x)) = P^*(z|x)$, which was instrumental for establishing transportability following equation (\ref{eq_transportability_post-treatment3}).} Interestingly though, $\mu$-transportability is feasible by combining information from both source domains. To see this, note that the post-intervention distribution in the target domain $\pi^*$ can be written as
\begin{align}
P^*(y|do(x)) &= \sum_z P^*(y|do(x),z) P^*(z|do(x)), \\[0.3cm]
&= \sum_z P^*(y|do(x),do(z)) P^*(z|do(x)), \label{eq_meta-trabsportability}
\end{align}
where the second line follows from rule 2 of do-calculus, since $(Z \independent Y | X)$ in $D_{\overline{X} \underline{Z}}$.\footnote{These do-calculus derivations are shown in detail, with corresponding subgraphs depicted next to it, in Appendix \ref{app_do-calculus3}.} Using this representation, each component can be shown to be individually transportable from one of the source domains. $P^*(z|do(x))$ is directly transportable from $\pi^a$, because $(S \independent Z)$ in $D_{\overline{X}}^{(a)}$. And $P^*(y|do(x),do(z))$ is directly transportable from $\pi^b$, since $(S \independent Y)$ in $D_{\overline{X,Z}}^{(b)}$. The individual components of equation (\ref{eq_meta-trabsportability}) can therefore be written as $P^*(z|do(x)) = P^{(a)}(z|do(x))$ and $P^*(y|do(x),do(z)) = P^{(b)}(y|do(x),do(z))$. This leads to the final transport formula
\begin{equation}
P^*(y|do(x)) = \sum_z P^{(b)}(y|do(x),do(z)) P^{(a)}(z|do(x)).
\end{equation}
In addition to demonstrating that multiple pairwise transportability is not a necessary condition for $\mu$-transportability, the example illustrates the superior inferential power obtained by combining multiple datasets over each individual dataset alone.

\citet{Bareinboim2013b} develop a complete algorithmic solution for deciding about $\mu$-transportability. The approach is further extended by \citet{Bareinboim2013c} who combine $\mu$-transportability with $\mathpzc{z}$-transportability, to allow for combining causal knowledge from multiple heterogeneous sources when only surrogate experiments on a subset $Z$ of variables in $\mathcal{D}$ are possible. This latter task is called \emph{mz-transportability} and can be automated by an algorithm that was proved to be complete by \citet{Bareinboim2014b}. 

In recent years, meta-analyses, which synthesize the results of several studies on a specific subject, are becoming increasingly important. Examples from economics can be found in \citet{Card2010}, \citet{Dehejia2015}, and \citet{Meager2019}. A drawback of standard meta-analytical approaches is, however, that they do not incorporate knowledge about domain heterogeneities related to causal mechanisms and background factors. Instead, they attempt to ``average out" differences across populations.\footnote{To the extent that these studies consider domain heterogeneity, this is done in a purely statistical fashion, without explicitly modelling structural differences across populations (\citealp{Dehejia2015, Meager2019}). This leaves open the question whether domains are actually structurally sufficiently similar for transportability to be feasible.} By contrast, the transportability techniques we have presented make it transparent how discrepancies between study results arise and how they can nonetheless be leveraged to identify a target query of interest in a principled and efficient manner. Moreover, they discipline the analyst to think carefully about the assumptions and shared mechanisms that allow extrapolation across domains to actually take place. 

Transportability theory thereby enables the research community to devise an effective strategy for leveraging the entire evidence base that exists related to a specific problem. Causal knowledge obtained by an individual experiment does not need to, and should not,  be regarded in isolation. Rather, it contributes to a larger body of empirical work that can be recombined to tackle entirely new policy problems, which were unimagined at the time of the original study. In combination with undergoing efforts to make more data sets openly available, transportability techniques thus bear the potential to save on discipline-wide data collection costs and to render causal inference a truly collective endeavor.\footnote{Other contributions to transportability theory have been made by \citet{Correa2019a}, who develop adjustment criteria for generalizing experimental findings in the presence of selection bias (see Section \ref{sec_selection}), and \citet{Lee2020}, who present a general treatment of transportability theory, unifying several of the techniques discussed in this section. 
Furthermore, these results have been extended to cover stochastic interventions, where the \emph{sigma calculus}, a generalization of the do-calculus, has been introduced (\citealp{correa2020calculus}), and  used to solve stochastic-transportability (\citealp{NEURIPS2020_7b497aa1}). More recently, these results were generalized for the case of transporting nested counterfactual from an arbitrary combination of observational and experimental distributions (\citealp{Correa2022}).}

   %%%%%%%%%%%%%%%%%%%%%%%%%%%%%%%%%%%%%%%%%%%%%%
\section{Conclusion}
\label{sec_conclusion}
%%%%%%%%%%%%%%%%%%%%%%%%%%%%%%%%%%%%%%%%%%%%%%

From the end of the 1980s onwards, the artificial intelligence field has developed an increased interest in causal inference (\citealp{Pearl1988, Pearl2009, Bareinboim2016, Pearl2018}). Causation is a fundamental concept in human thinking and structures the way in which we interact with our environment (\citealp{Sloman2005}). A human-like AI, therefore, needs to possess an internal representation of causality in order to mimic human behaviour and communicate with us in a meaningful way (\citealp{Pearl2018,bareinboim2020pch}). Tremendous progress over the last three decades has led to the development of a powerful causal inference engine, which puts an artificial learner into the position to acquire and combine causal knowledge from many diverse sources in its surroundings. In particular, several important contributions to the literature in recent years have made this engine more robust, general, and practical, by expanding its applicability to various different data collection and knowledge contexts (\citealp{Bareinboim2016,bareinboim2020pch}).

We are convinced that the causal inference and data fusion techniques we have discussed in this paper also have a lot to offer to econometricians. Until today, the possibilities to completely automate the identification task, which is a necessary ingredient for causal machine learning, remain largely unexplored in econometrics. The applications of do-calculus we have discussed only require the analyst to provide a model of the economic context under study and a description of the available data, the rest can be handled automatically by an algorithm.\footnote{Up to a certain extent, directed acyclic graphs can also be learned from observational data. Respective techniques rely on the testable implications of DAGs that were discussed in Section \ref{sec_preliminaries}  to find an equivalence class of models that is compatible with the d-separation relations in the data. The interested reader is referred to the literature on ``causal structure learning" and ``causal discovery" in the AI field (\citealp{Spirtes2001,Pearl2009,Peters2017}). Automation of the identification task in these settings has also gained traction recently  (\citealp{zhang2006causal,perkovic2017complete,jaber2018graphical,jaber2018causal, jaber2019causalcomplete,Jaber2022}).}
Moreover, graphical representations of structural causal models do not require the learner -- whether artificial or human -- to impose any distributional or functional form restrictions on the underlying causal mechanisms under study. The approach remains fully nonparametric and thus naturally incorporates, e.g., treatment effect heterogeneity. At the same time, crucial identification conditions, such as conditional independence, are derived from the properties of the underlying structural model, rather than being assumed to hold \emph{a priori}. Causal graphical models thus combine the flexibility and accessibility of potential outcomes with the analytical rigor of structural econometrics (\citealp{Rust1987,Keane1997,Heckman2007}). These properties are of great value for applied empirical work. Economists should therefore feel encouraged to engage in a productive exchange with AI researchers for mutual benefit from the numerous useful tools for causal inference developed in both disciplines.

	% acknowledgements
	
	\section*{Acknowledgements}
	We would like to thank Jaap Abbring and two anonymous referees for excellent editorial guidance and useful comments. Furthermore, we are grateful to Victor Chernozhukov, Carlos Cinelli, Juan Correa, Felix Elwert, Guido Imbens, Beyers Louw, Judea Pearl, attendees at EEA-ESEM 2019, VHB Annual Meeting 2020, and seminar participants at Booking.com, Boston University, Duisburg-Essen, Harvard University, Hebrew University, Humboldt University, ifo Institute, JADS, KU Leuven, LMU Munich, Maastricht University, MIT, RWTH Aachen, SEACEN Centre, Stanford GSB, Technion, TU Dortmund, Vinted, and WZB Berlin for their comments and suggestions.

    % references

	\bibliographystyle{chicago}
    \bibliography{fusion_references_EJ.bib}

    % appendix (put online supplement in a separate file)
	
	%%%%%%%%%%%%%%%%%%%%%%%%%%%%%%%%%%%%%%%%%%%%%%
\section*{Appendix}
\label{app_do-calculus}
%%%%%%%%%%%%%%%%%%%%%%%%%%%%%%%%%%%%%%%%%%%%%%
\renewcommand{\theequation}{A.\arabic{equation}}
\renewcommand{\thesection}{A}
\setcounter{equation}{0}

This appendix shows step-by-step solutions for the do-calculus derivations discussed in the main text. For illustration purposes, subgraphs used in the respective steps are placed alongside.

%%%%%%%%%%%%%%%%%%%%%%%%%%%%%
\subsection{Selection bias example (Section \ref{sec_selection}, Figure \ref{fig_selection_algorithm})}
\label{app_do-calculus2}
%%%%%%%%%%%%%%%%%%%%%%%%%%%%%

\begin{figure}[htp]
	\centering
	\subcaptionbox*{}[.48\linewidth]{
		\begin{tikzpicture}[>=triangle 45, font=\scriptsize]
		\node[draw] at (0.3,2.2) {\normalsize $G$};
		\node[fill,circle,inner sep=0pt,minimum size=5pt,label={left:{Z}}] (Z) at (0,0) {};
		\node[fill,circle,inner sep=0pt,minimum size=5pt,label={above:{W}}] (W) at (1,1.2) {};
		\node[fill,circle,inner sep=0pt,minimum size=5pt,label={below:{X}}] (X) at (2,0) {};
		\node[fill,circle,inner sep=0pt,minimum size=5pt,label={right:{Y}}] (Y) at (4,0) {};
		\node[draw,circle,inner sep=0pt,double,double distance=0.3mm,minimum size=5pt,label={above:{S}}] (S) at (2.8,0.8) {};
		\draw[->,shorten >= 1pt] (X)--(Y);
		\draw[->,shorten >= 1pt] (Z)--(X);
		\draw[->,shorten >= 1pt] (W)--(X);
		\draw[->,shorten >= 1pt] (W)--(Z);
		\draw[->,shorten >= 2pt] (W)--(S);
		\draw[<->,dashed,shorten >= 1pt] (Z) to[bend right=45] (Y);
		\end{tikzpicture}
	}
	\subcaptionbox*{}[.48\linewidth]{
		\begin{tikzpicture}[>=triangle 45, font=\scriptsize]
		\node[draw] at (0.3,2.3) {\normalsize $G_{\overline{X}}$};
		\node[fill,circle,inner sep=0pt,minimum size=5pt,label={left:{Z}}] (Z) at (0,0) {};
		\node[fill,circle,inner sep=0pt,minimum size=5pt,label={above:{W}}] (W) at (1,1.2) {};
		\node[fill,circle,inner sep=0pt,minimum size=5pt,label={below:{X}}] (X) at (2,0) {};
		\node[fill,circle,inner sep=0pt,minimum size=5pt,label={right:{Y}}] (Y) at (4,0) {};
		\node[draw,circle,inner sep=0pt,double,double distance=0.3mm,minimum size=5pt,label={above:{S}}] (S) at (2.8,0.8) {};
		\draw[->,shorten >= 1pt] (X)--(Y);
		\draw[->,shorten >= 1pt] (W)--(Z);
		\draw[->,shorten >= 2pt] (W)--(S);
		\draw[<->,dashed,shorten >= 1pt] (Z) to[bend right=45] (Y);
		\end{tikzpicture}
	} \\
	\subcaptionbox*{}[.48\linewidth]{
		\begin{tikzpicture}[>=triangle 45, font=\scriptsize]
		\node[draw] at (0.3,2.3) {\normalsize $G_{\underline{X}}$};
		\node[fill,circle,inner sep=0pt,minimum size=5pt,label={left:{Z}}] (Z) at (0,0) {};
		\node[fill,circle,inner sep=0pt,minimum size=5pt,label={above:{W}}] (W) at (1,1.2) {};
		\node[fill,circle,inner sep=0pt,minimum size=5pt,label={below:{X}}] (X) at (2,0) {};
		\node[fill,circle,inner sep=0pt,minimum size=5pt,label={right:{Y}}] (Y) at (4,0) {};
		\node[draw,circle,inner sep=0pt,double,double distance=0.3mm,minimum size=5pt,label={above:{S}}] (S) at (2.8,0.8) {};
		\draw[->,shorten >= 1pt] (Z)--(X);
		\draw[->,shorten >= 1pt] (W)--(X);
		\draw[->,shorten >= 1pt] (W)--(Z);
		\draw[->,shorten >= 2pt] (W)--(S);
		\draw[<->,dashed,shorten >= 1pt] (Z) to[bend right=45] (Y);
		\end{tikzpicture}
	}
	\subcaptionbox*{}[.48\linewidth]{
		\begin{tikzpicture}[>=triangle 45, font=\scriptsize]
		\node[draw] at (0.3,2.3) {\normalsize $G_{\overline{X(W)}}$};
		\node[fill,circle,inner sep=0pt,minimum size=5pt,label={left:{Z}}] (Z) at (0,0) {};
		\node[fill,circle,inner sep=0pt,minimum size=5pt,label={above:{W}}] (W) at (1,1.2) {};
		\node[fill,circle,inner sep=0pt,minimum size=5pt,label={below:{X}}] (X) at (2,0) {};
		\node[fill,circle,inner sep=0pt,minimum size=5pt,label={right:{Y}}] (Y) at (4,0) {};
		\node[draw,circle,inner sep=0pt,double,double distance=0.3mm,minimum size=5pt,label={above:{S}}] (S) at (2.8,0.8) {};
		\draw[->,shorten >= 1pt] (X)--(Y);
		\draw[->,shorten >= 1pt] (W)--(Z);
		\draw[->,shorten >= 2pt] (W)--(S);
		\draw[<->,dashed,shorten >= 1pt] (Z) to[bend right=45] (Y);
		\end{tikzpicture}
	}
\end{figure}

Consider the causal effect of $X$ on $Y$ in graph $G$. In graph $G_{\overline{X}}$, $Z$ is a collider on the path connecting $S$ and $W$ with $Y$. Therefore, $(S,W \independent Y)_{G_{\overline{X}}}$, and by the first rule of do-calculus it holds that
\begin{align*}
P(y|do(x)) &= P(y|do(x), w, S=1),  \\[0.3cm]
&= \sum_z P(y|do(x), z, w, S=1) P(z|do(x), w, S=1).
\end{align*}
Moreover, because $(Y \independent X | W,Z, S)$ in $G_{\underline{X}}$, rule 2 of do-calculus applies to the first factor, which leads to 
\begin{equation*}
P(y|do(x)) = \sum_z P(y|x, z, w, S=1) P(z|do(x), w, S=1).
\end{equation*}
Finally, notice that there are no open paths between $X$ and $Z$ in $G_{\overline{X(W)}}$ (which is equivalent to $G_{\overline{X}}$, as $X$ is not an ancestor of $W$). Thus, since $(Z \independent X | W, S)_{G_{\overline{X(W)}}}$ (this independence holds both conditional and unconditional), rule 3 of do-calculus can be applied to the second term, such that
\begin{equation*}
P(y|do(x)) = \sum_z P(y|x, z, w, S=1) P(z| w, S=1).
\end{equation*}

\hfill$\square$\\

%%%%%%%%%%%%%%%%%%%%%%%%%%%%%
\subsection{\textit{M}-Transportability example (Section \ref{sec_m-transportability}, Figure \ref{fig9})}
\label{app_do-calculus3}
%%%%%%%%%%%%%%%%%%%%%%%%%%%%%

\begin{figure}[htp]
	\centering
	\subcaptionbox*{}[.32\linewidth]{
		\begin{tikzpicture}[>=triangle 45, font=\scriptsize, scale=0.8]
		\node[draw] at (0.3,2) {\normalsize $D$};
		\node[fill,circle,inner sep=0pt,minimum size=5pt,label={[xshift=-2]below:{X}}] (X) at (0,0) {};
		\node[fill,circle,inner sep=0pt,minimum size=5pt,label={[xshift=2]below:{Y}}] (Y) at (4,0) {};
		\node[fill,circle,inner sep=0pt,minimum size=5pt,label={below:{Z}}] (Z) at (2,0) {};
		\draw[->,shorten >= 1pt] (X)--(Z);
		\draw[->,shorten >= 1pt] (Z)--(Y);
		\draw[->,shorten >= 1pt] (X) to[bend right=45] (Y);
		\draw[<->,dashed,shorten >= 1pt] (X) to[bend left=75] (Y);
		\draw[<->,dashed,shorten >= 1pt] (X) to[bend left=45] (Z);
		\end{tikzpicture}
	}
	\subcaptionbox*{}[.32\linewidth]{
		\begin{tikzpicture}[>=triangle 45, font=\scriptsize, scale=0.8]
		\node[draw] at (0.3,2) {\normalsize $D_{\overline{X} \underline{Z}}$};
		\node[fill,circle,inner sep=0pt,minimum size=5pt,label={[xshift=-2]below:{X}}] (X) at (0,0) {};
		\node[fill,circle,inner sep=0pt,minimum size=5pt,label={[xshift=2]below:{Y}}] (Y) at (4,0) {};
		\node[fill,circle,inner sep=0pt,minimum size=5pt,label={below:{Z}}] (Z) at (2,0) {};
		\draw[->,shorten >= 1pt] (X)--(Z);
		\draw[->,shorten >= 1pt] (X) to[bend right=45] (Y);
		\end{tikzpicture}
	}
	\subcaptionbox*{}[.32\linewidth]{
		\begin{tikzpicture}[>=triangle 45, font=\scriptsize, scale=0.8]
		\node[draw] at (0.3,2.1) {\normalsize $D^{(a)}$};
		\node[fill,rectangle,inner sep=0pt,minimum size=5pt,label={right:{$S_1$}}] (S1) at (0,1.2) {};
		\node[fill,rectangle,inner sep=0pt,minimum size=5pt,label={left:{$S_2$}}] (S2) at (4,1.2) {};
		\node[fill,circle,inner sep=0pt,minimum size=5pt,label={[xshift=-2]below:{X}}] (X) at (0,0) {};
		\node[fill,circle,inner sep=0pt,minimum size=5pt,label={[xshift=2]below:{Y}}] (Y) at (4,0) {};
		\node[fill,circle,inner sep=0pt,minimum size=5pt,label={below:{Z}}] (Z) at (2,0) {};
		\draw[->,shorten >= 1pt] (X)--(Z);
		\draw[->,shorten >= 1pt] (Z)--(Y);
		\draw[->,shorten >= 1pt] (S1)--(X);
		\draw[->,shorten >= 1pt] (S2)--(Y);
		\draw[->,shorten >= 1pt] (X) to[bend right=45] (Y);
		\draw[<->,dashed,shorten >= 1pt] (X) to[bend left=75] (Y);
		\draw[<->,dashed,shorten >= 1pt] (X) to[bend left=45] (Z);
		\end{tikzpicture}
	} \\
	\subcaptionbox*{}[.32\linewidth]{
		\begin{tikzpicture}[>=triangle 45, font=\scriptsize, scale=0.8]
		\node[draw] at (0.3,2) {\normalsize $D^{(a)}_{\overline{X}}$};
		\node[fill,rectangle,inner sep=0pt,minimum size=5pt,label={right:{$S_1$}}] (S1) at (0,1.2) {};
		\node[fill,rectangle,inner sep=0pt,minimum size=5pt,label={left:{$S_2$}}] (S2) at (4,1.2) {};
		\node[fill,circle,inner sep=0pt,minimum size=5pt,label={[xshift=-2]below:{X}}] (X) at (0,0) {};
		\node[fill,circle,inner sep=0pt,minimum size=5pt,label={[xshift=2]below:{Y}}] (Y) at (4,0) {};
		\node[fill,circle,inner sep=0pt,minimum size=5pt,label={below:{Z}}] (Z) at (2,0) {};
		\draw[->,shorten >= 1pt] (X)--(Z);
		\draw[->,shorten >= 1pt] (Z)--(Y);
		\draw[->,shorten >= 1pt] (S2)--(Y);
		\draw[->,shorten >= 1pt] (X) to[bend right=45] (Y);
		\end{tikzpicture}
	}
	\subcaptionbox*{}[.32\linewidth]{
		\begin{tikzpicture}[>=triangle 45, font=\scriptsize, scale=0.8]
		\node[draw] at (0.3,2) {\normalsize $D^{(b)}$};
		\node[fill,rectangle,inner sep=0pt,minimum size=5pt,label={right:{$S_3$}}] (S3) at (0,1.2) {};
		\node[fill,rectangle,inner sep=0pt,minimum size=5pt,label={right:{$S_4$}}] (S4) at (2,0.9) {};
		\node[fill,circle,inner sep=0pt,minimum size=5pt,label={[xshift=-2]below:{X}}] (X) at (0,0) {};
		\node[fill,circle,inner sep=0pt,minimum size=5pt,label={[xshift=2]below:{Y}}] (Y) at (4,0) {};
		\node[fill,circle,inner sep=0pt,minimum size=5pt,label={below:{Z}}] (Z) at (2,0) {};
		\draw[->,shorten >= 1pt] (X)--(Z);
		\draw[->,shorten >= 1pt] (Z)--(Y);
		\draw[->,shorten >= 1pt] (S3)--(X);
		\draw[->,shorten >= 1pt] (S4)--(Z);
		\draw[->,shorten >= 1pt] (X) to[bend right=45] (Y);
		\draw[<->,dashed,shorten >= 1pt] (X) to[bend left=75] (Y);
		\draw[<->,dashed,shorten >= 1pt] (X) to[bend left=45] (Z);
		\end{tikzpicture}
	}
	\subcaptionbox*{}[.32\linewidth]{
		\begin{tikzpicture}[>=triangle 45, font=\scriptsize, scale=0.8]
		\node[draw] at (0.3,2.4) {\normalsize $D^{(b)}_{\overline{X,Z}}$};
		\node[fill,rectangle,inner sep=0pt,minimum size=5pt,label={right:{$S_3$}}] (S3) at (0,0.9) {};
		\node[fill,rectangle,inner sep=0pt,minimum size=5pt,label={right:{$S_4$}}] (S4) at (2,0.9) {};
		\node[fill,circle,inner sep=0pt,minimum size=5pt,label={[xshift=-2]below:{X}}] (X) at (0,0) {};
		\node[fill,circle,inner sep=0pt,minimum size=5pt,label={[xshift=2]below:{Y}}] (Y) at (4,0) {};
		\node[fill,circle,inner sep=0pt,minimum size=5pt,label={below:{Z}}] (Z) at (2,0) {};
		\draw[->,shorten >= 1pt] (Z)--(Y);
		\end{tikzpicture}
	}
\end{figure}

Consider the causal effect of $X$ on $Y$ in graph $D$, in target domain $\pi^*$: 
\begin{equation*}
P^*(y|do(x)).
\end{equation*}
Note that $X$ d-separates $Z$ and $Y$ in $D_{\overline{X} \underline{Z}}$. Thus, since $(Z \independent Y | X)_{D_{\overline{X} \underline{Z}}}$, it follows from rule 2 of do-calculus that
\begin{align*}
P^*(y|do(x)) &= \sum_z P^*(y|do(x),z) P^*(z|do(x)), \\[0.3cm]
&= \sum_z P^*(y|do(x),do(z)) P^*(z|do(x)).
\end{align*}
Let the selection diagrams for the two source domains $\pi^a$ and $\pi^b$ be given by $D^{(a)}$ and $D^{(b)}$, respectively. 
Note that $(S_1, S_2 \independent Z)$ in $D_{\overline{X}}^{(a)}$, therefore, $P^*(z|do(x))$ is directly transportable from $\pi^a$ as
\begin{equation*}
P^*(z|do(x)) = P^{(a)}(z|do(x)).
\end{equation*}
Furthermore, since $(S_3, S_4 \independent Y)$ in $D_{\overline{X,Z}}^{(b)}$, $P^*(y|do(x),do(z))$ is directly transportable from $\pi^b$
\begin{equation*}
P^*(y|do(x),do(z)) = P^{(b)}(y|do(x),do(z)).
\end{equation*}
Combining the two expressions leads to the final transport formula
\begin{equation*}
P^*(y|do(x)) = \sum_z P^{(b)}(y|do(x),do(z)) P^{(a)}(z|do(x)).
\end{equation*}

\hfill$\square$\\

\end{document}